\newif\ifOneCol

\ifOneCol
	\documentclass[draftcls,onecolumn,11pt]{IEEEtran}
\else
	\documentclass[twocolumn,10pt]{IEEEtran}
\fi

\usepackage[cmex10]{amsmath}
\usepackage{amsfonts}
\usepackage{verbatim}
\usepackage{algpseudocode}

\usepackage{cite}

\ifCLASSINFOpdf
  \usepackage[pdftex]{graphicx}
  \graphicspath{{graphics/}}
  \DeclareGraphicsExtensions{.pdf,.jpeg,.png}
\else
  \usepackage[dvips]{graphicx}
  \graphicspath{{graphics/}}
  \DeclareGraphicsExtensions{.eps}
\fi

\usepackage{array}

\usepackage{mdwmath}
\usepackage{mdwtab}
\DeclareMathOperator{\erf}{erf}

\begin{document}
	
\ifOneCol
\sloppy 
\fi

\bibliographystyle{IEEEtran}
%
\title{Optimal Receiver Design for Diffusive Molecular Communication
With Flow and Additive Noise}


\author{Adam Noel, \IEEEmembership{Student Member, IEEE}, Karen C. Cheung, and
	Robert Schober,
	\IEEEmembership{Fellow, IEEE}
	\thanks{Manuscript received July 31, 2013; revised June 18, 2014; accepted
		July 7, 2014. This work was supported by the Natural
		Sciences and Engineering Research Council of Canada, and a Walter C. Sumner
		Memorial Fellowship.}
	\thanks{The authors are with the Department of Electrical and Computer Engineering, University of British Columbia, Vancouver, BC, Canada,	V6T 1Z4 (email: \{adamn, kcheung, rschober\}@ece.ubc.ca). R. Schober is also with the Institute for Digital Communication, Friedrich-Alexander-Universit\"{a}t Erlangen-N\"{u}rnberg (FAU), Erlangen, Germany (email: schober@lnt.de).}}


\newcommand{\dbydt}[1]{\frac{d#1}{dt}}
\newcommand{\pbypx}[2]{\frac{\partial #1}{\partial #2}}
\newcommand{\psbypxs}[2]{\frac{\partial^2 #1}{\partial {#2}^2}}
\newcommand{\dbydtc}[1]{\dbydt{\conc{#1}}}
\newcommand{\thev}{\theta_v}
\newcommand{\thevi}[1]{\theta_{v#1}}
\newcommand{\theh}{\theta_h}
\newcommand{\thehi}[1]{\theta_{h#1}}
\newcommand{\x}{x}
\newcommand{\y}{y}
\newcommand{\z}{z}
\newcommand{\rad}[1]{\vec{r}_{#1}}
\newcommand{\radmag}[1]{|\rad{#1}|}

\newcommand{\kth}[1]{k_{#1}}
\newcommand{\km}{K_M}
\newcommand{\vm}{v_{max}}
\newcommand{\conc}[1]{[#1]}
\newcommand{\conco}[1]{[#1]_0}
\newcommand{\C}{C}
\newcommand{\Cx}[1]{C_{#1}}
\newcommand{\CxFun}[3]{C_{#1}(#2,#3)}
\newcommand{\Cobs}{C_{obs}}
\newcommand{\Nobs}{{\Nx{\A}}_{obs}}
\newcommand{\Nobst}[1]{\Nobs\!\left(#1\right)}
\newcommand{\Nobsn}[1]{\Nobs\left[#1\right]}
\newcommand{\Nobsavgt}{\overline{{\Nx{\A}}_{obs}}(t)}
\newcommand{\Nobsavg}[1]{\overline{{\Nx{\A}}_{obs}}\left(#1\right)}
\newcommand{\Nobsavgmax}{\overline{{\Nx{\A}}_{max}}}
\newcommand{\Nnoisetavg}[1]{\overline{{\Nx{\A}}_{noise}}\left(#1\right)}
\newcommand{\Nnoiset}[1]{{\Nx{\A}}_{noise}\left(#1\right)}
\newcommand{\Ntxt}[1]{{\Nx{\A}}_{TX}\left(#1\right)}
\newcommand{\Ntxtavg}[1]{\overline{{\Nx{\A}}_{TX}}\left(#1\right)}
\newcommand{\Cgen}{C_A(r, t)}
\newcommand{\radbind}{r_B}

\newcommand{\M}{M}
\newcommand{\smM}{m}
\newcommand{\A}{A}
\newcommand{\X}{S}
\newcommand{\vx}[1]{v_{#1}}
\newcommand{\metre}{\textnormal{m}}
\newcommand{\second}{\textnormal{s}}
\newcommand{\molecule}{\textnormal{molecule}}
\newcommand{\bound}{\textnormal{bound}}
\newcommand{\argmax}{\operatornamewithlimits{argmax}}
\newcommand{\Dx}[1]{D_{#1}}
\newcommand{\Nx}[1]{N_{#1}}
\newcommand{\Nemit}{\Nx{{\A_{EM}}}}
\newcommand{\Da}{D_\A}
\newcommand{\En}{E}
\newcommand{\en}{e}
\newcommand{\Ne}{\Nx{\En}}
\newcommand{\De}{D_\En}
\newcommand{\EA}{EA}
\newcommand{\ea}{ea}
\newcommand{\Nint}{\Nx{\EA}}
\newcommand{\Di}{D_{\EA}}
\newcommand{\Etot}{\En_{Tot}}
\newcommand{\stepl}{r_{rms}}
\newcommand{\AP}{A_P}
\newcommand{\Ri}[1]{R_{#1}}
\newcommand{\ro}{r_0}
\newcommand{\rone}{r_1}
\newcommand{\visc}{\eta}
\newcommand{\bolt}{\kth{B}}
\newcommand{\temp}{T}
\newcommand{\T}{T_{int}}
\newcommand{\Vobs}{V_{obs}}
\newcommand{\robs}{r_{obs}}
\newcommand{\Ve}{V_{enz}}
\newcommand{\tint}{\delta t}
\newcommand{\tmax}{t_{max}}
\newcommand{\Cobsfrac}{\alpha}
\newcommand{\dist}{L}
\newcommand{\DMLSA}{a}
\newcommand{\DMLSt}[1]{t_{#1}^\star}
\newcommand{\DMLSx}{x^\star}
\newcommand{\DMLSy}{y^\star}
\newcommand{\DMLSz}{z^\star}
\newcommand{\DMLSr}{r_{obs}^\star}
\newcommand{\DMLSrad}[1]{\rad{#1}^\star}
\newcommand{\DMLSradmag}[1]{|\DMLSrad{#1}|}
\newcommand{\DMLSC}[1]{\Cx{#1}^\star}
\newcommand{\DMLSCxFun}[3]{{\DMLSC{#1}}(#2,#3)}
\newcommand{\DMLSc}[1]{\gamma_{#1}}
\newcommand{\DMLSV}{\Vobs^\star}
\newcommand{\DMLSNA}{\overline{{\Nx{\DMLSA}}_{obs}^\star}(t)}
\newcommand{\DMLSNAb}{\overline{{\Nx{\DMLSA}}_{obs}^\star}(\DMLSt{B})}
\newcommand{\DMLSNAmax}{{\overline{{\Nx{\DMLSA}}_{max}^\star}}}
\newcommand{\DMLStmax}[1]{{\DMLSt{#1}}_{,max}}
\newcommand{\DMLSdim}{\mathcal{D}}
\newcommand{\DMLSthreshInt}{\alpha^\star}

\newcommand{\data}[1]{W\left[#1\right]}
\newcommand{\dataSeq}{\mathbf{W}}
\newcommand{\dataSet}{\mathcal{W}}
\newcommand{\dataObs}[1]{\hat{W}\left[#1\right]}
\newcommand{\numX}[2]{n_{#1}\left(#2\right)}
\newcommand{\thresh}{\xi}
\newcommand{\poissBar}{\Big|_\textnormal{Poiss}}
\newcommand{\gaussBar}{\Big|_\textnormal{Gauss}}
\newcommand{\eqBar}[2]{\Big|_{#1 = #2}}
\newcommand{\Pobs}{P_{obs}}
\newcommand{\Pobsx}[1]{P_{obs}\left(#1\right)}
\newcommand{\Pone}{P_1}
\newcommand{\Pzero}{P_0}
\newcommand{\Pe}[1]{P_e\left[#1\right]}
\newcommand{\Peavg}[1]{\overline{P}_e\left[#1\right]}
\newcommand{\threshInterval}{\alpha}
\newcommand{\pstay}[1]{P_{stay}\left(#1\right)}
\newcommand{\pleave}[1]{P_{leave}\left(#1\right)}
\newcommand{\parrive}[1]{P_{arr}\left(#1\right)}

\newcommand{\VAmem}{F}
\newcommand{\VAstate}{f}
\newcommand{\VAdataObs}[2]{\hat{W}_{\VAstate_{#2}}\left[#1\right]}
\newcommand{\VAcurLL}[2]{\Phi_{\VAstate_{#2}}\left[#1\right]}
\newcommand{\VAcumLL}[2]{L_{\VAstate_{#2}}\left[#1\right]}

\newcommand{\weight}[1]{w_{#1}}

\newcommand{\fof}[1]{f\left(#1\right)}
\newcommand{\floor}[1]{\lfloor#1\rfloor}
\newcommand{\lam}[1]{W\left(#1\right)}
\newcommand{\EXP}[1]{\exp\left(#1\right)}
\newcommand{\ERF}[1]{\erf\left(#1\right)}
\newcommand{\SIN}[1]{\sin\left(#1\right)}
\newcommand{\SINH}[1]{\sinh\left(#1\right)}
\newcommand{\COS}[1]{\cos\left(#1\right)}
\newcommand{\COSH}[1]{\cosh\left(#1\right)}
\newcommand{\Ix}[2]{I_{#1}\!\left(#2\right)}
\newcommand{\Jx}[2]{J_{#1}\!\left(#2\right)}
\newcommand{\E}[1]{E\left[#1\right]}
\newcommand{\GamFcn}[1]{\Gamma\!\left(#1\right)}
\newcommand{\mean}[1]{\mu_{#1}}
\newcommand{\var}[1]{\sigma_{#1}^2}

\newcommand{\B}[1]{B_{#1}}
\newcommand{\w}{w}
\newcommand{\n}{n}
\newcommand{\gx}[1]{g\left(#1\right)}
\newcommand{\hx}[1]{h\left(#1\right)}
\newcommand{\tx}[1]{t\left(#1\right)}
\newcommand{\ux}[1]{u\left(#1\right)}
\newcommand{\deltObs}{t_{o}}
\newcommand{\sx}[1]{s_{#1}}

\newcommand{\new}[1]{\textbf{#1}}
\newcommand{\ISI}{ISI}
\newcommand{\DDFSE}{DDFSE}
\newcommand{\PDF}{PDF}
\newcommand{\CDF}{CDF}
\newcommand{\AWGN}{AWGN}

\newtheorem{theorem}{Theorem}


\newcommand{\figOne}[1]{
	\begin{figure}[#1]
	\centering
	\includegraphics[width=\linewidth]
	{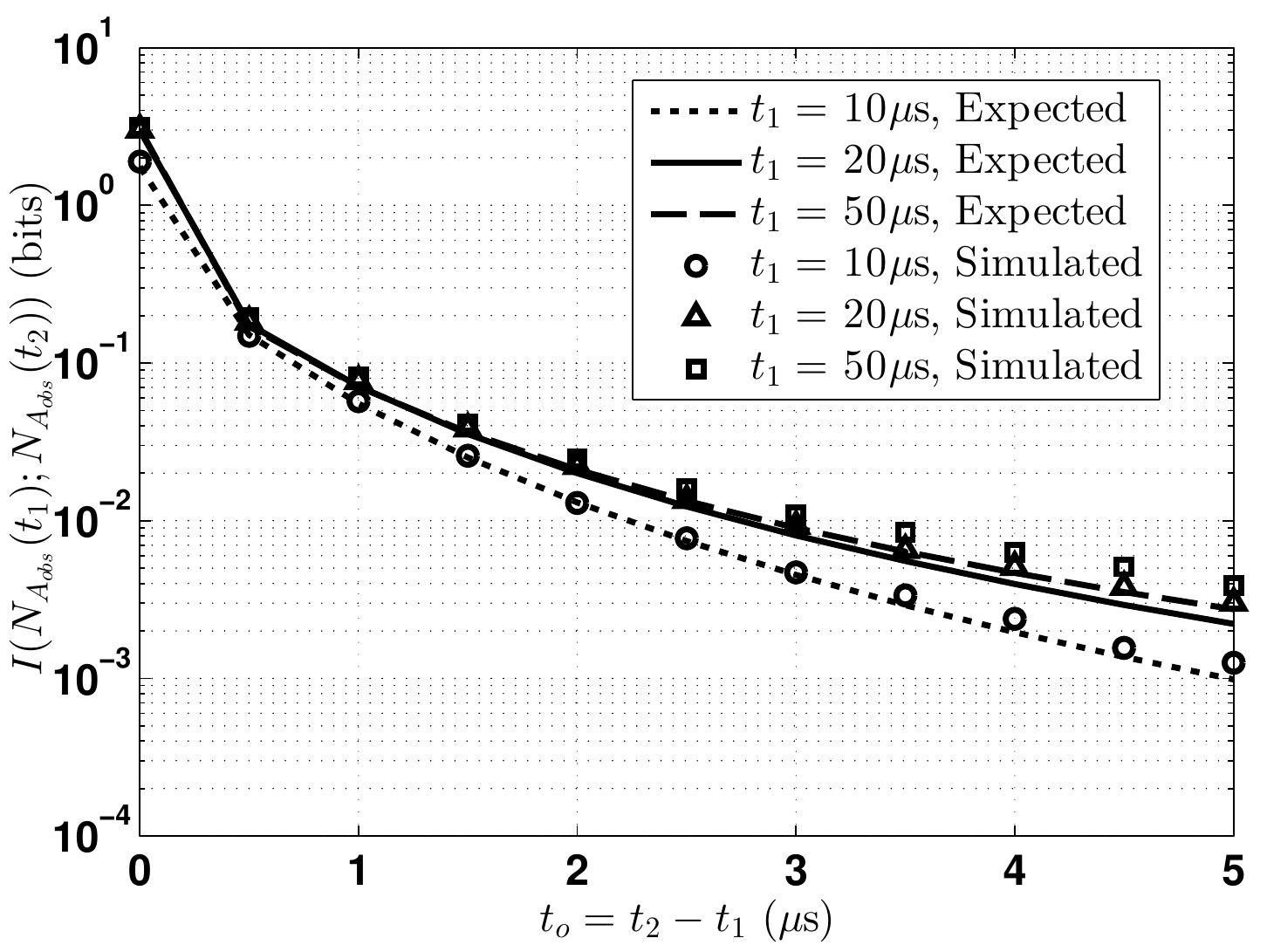}
	\caption{The mutual information in bits measured as a function
	of $\deltObs$.}
	\label{fig1}
	\end{figure}
}

\newcommand{\figThree}[1]{
	\begin{figure}[#1]
	\centering
	\includegraphics[width=\linewidth]
	{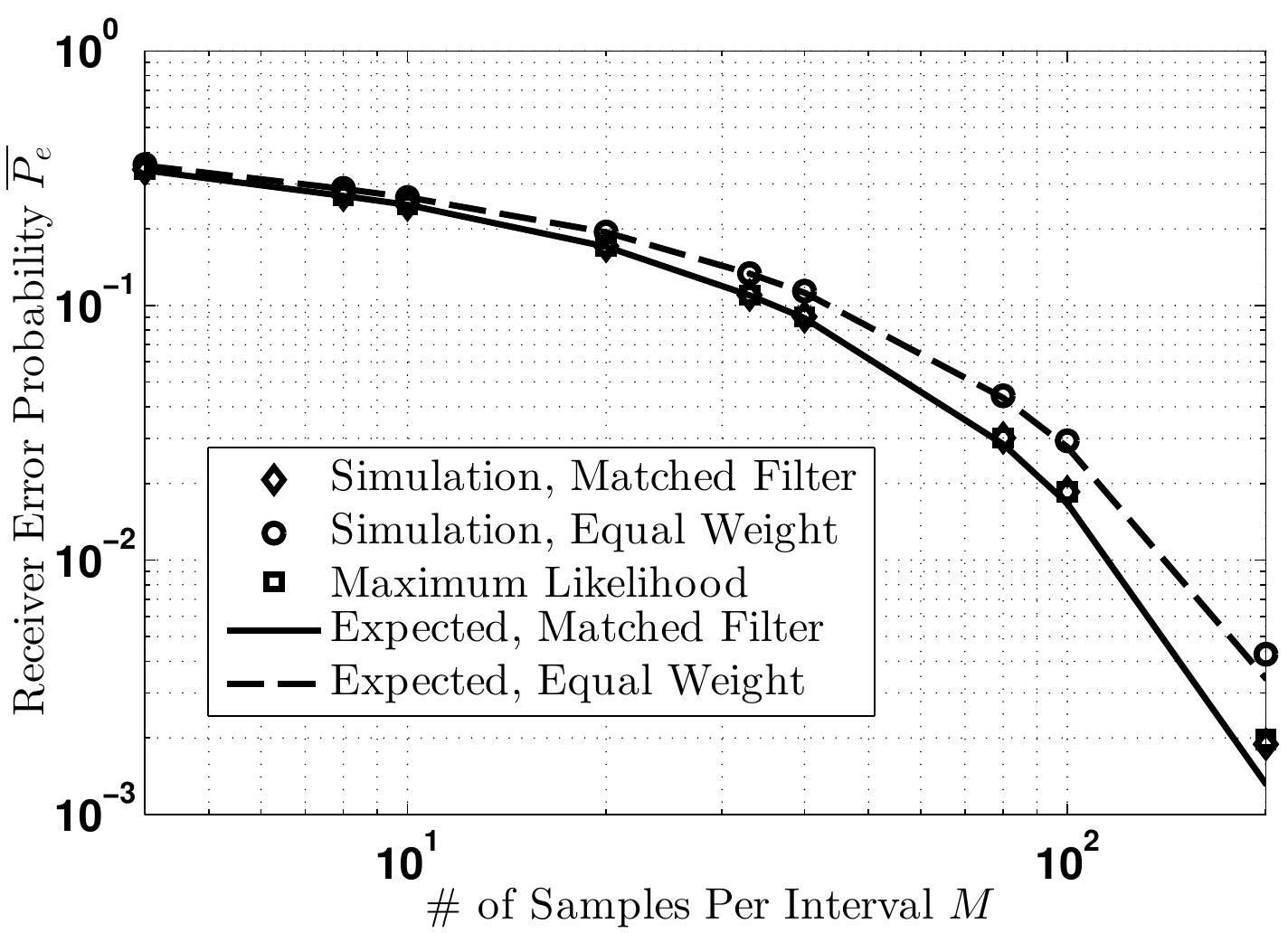}
	\caption{Expected error probability as a function of $\M$
	when there is no \ISI,
	$\Nnoisetavg{t} = 50$, and $\T = 200\,\mu\second$. The performance
	of the matched filter detector is equivalent to that of the
	maximum likelihood detector.}
	\label{fig3}
	\end{figure}
}

\newcommand{\figThreePointFive}[1]{
	\begin{figure}[#1]
		\centering
		\includegraphics[width=\linewidth]
		{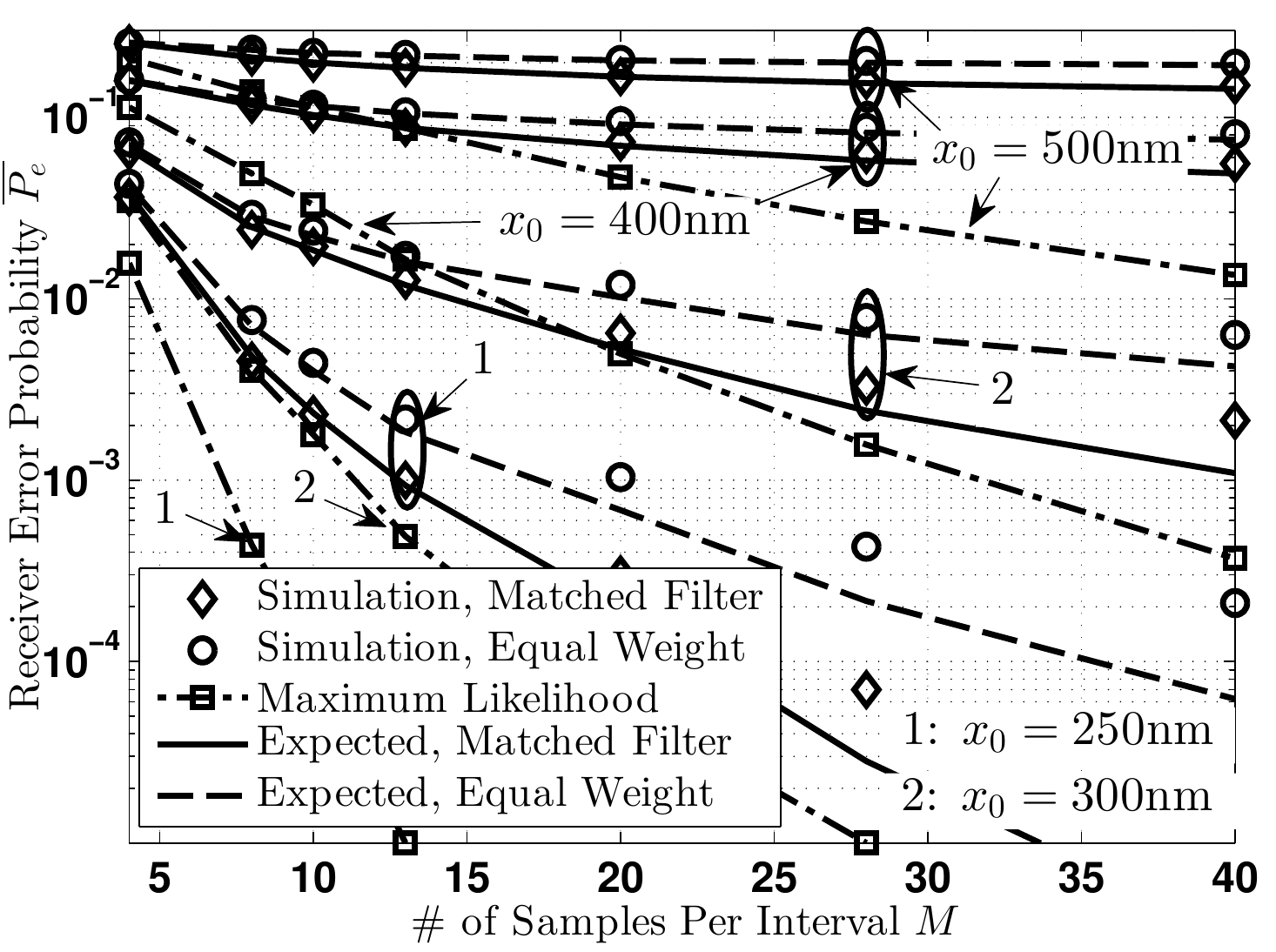}
		\caption{Receiver error probability as a function of $\M$
			when \ISI\, is included, $\T = 200\,\mu\second$,
			and the distance $\x_0$ to the receiver is varied.}
		\label{fig3_5}
	\end{figure}
}

\newcommand{\figFour}[1]{
	\begin{figure}[#1]
	\centering
	\includegraphics[width=\linewidth]
	{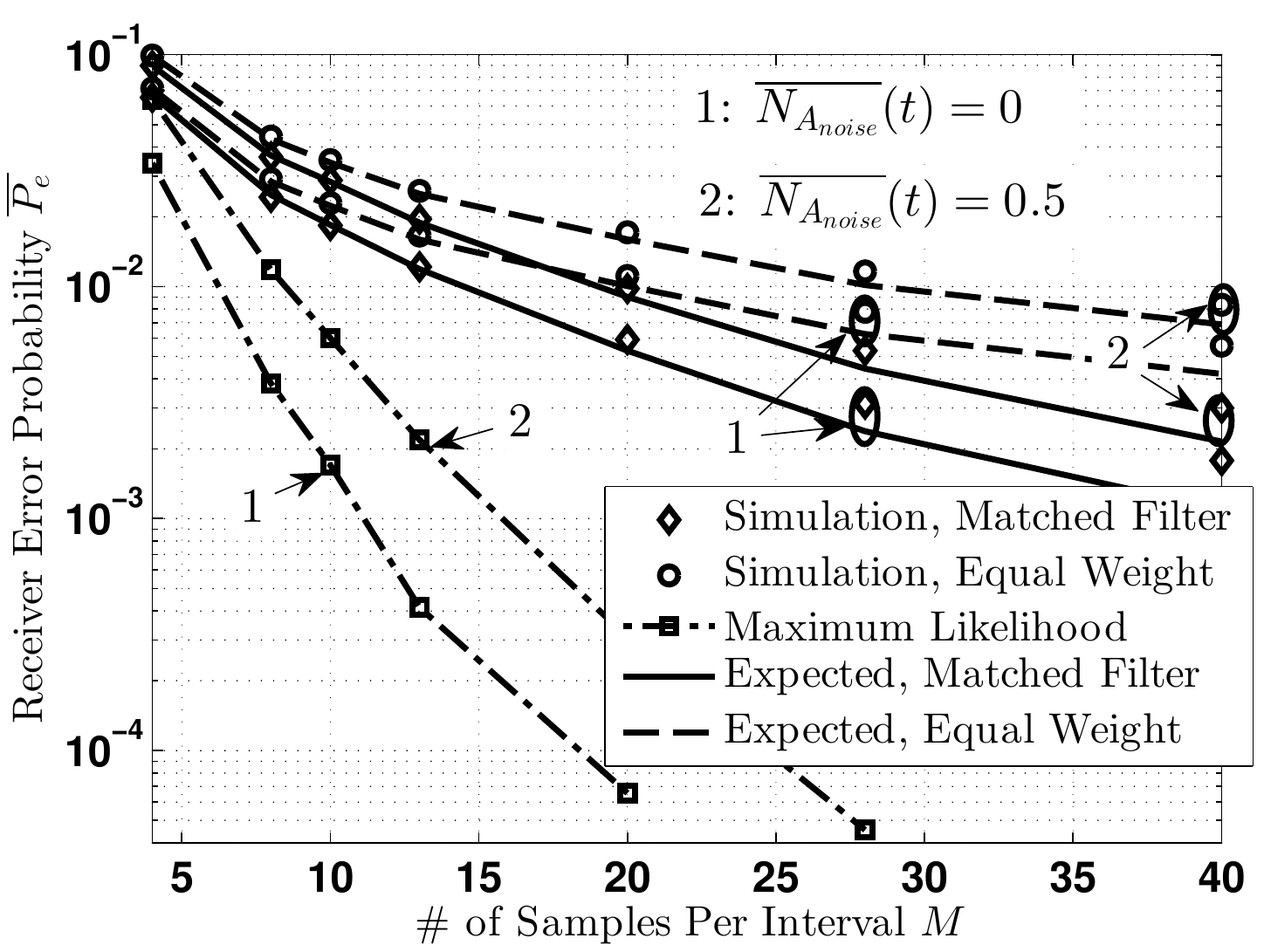}
	\caption{Receiver error probability as a function of $\M$
	when \ISI\, is included, $\T = 200\,\mu\second$,
	and $\Nnoisetavg{t} = 0$ or $0.5$.}
	\label{fig4}
	\end{figure}
}

\newcommand{\figFive}[1]{
	\begin{figure}[#1]
	\centering
	\includegraphics[width=\linewidth]
	{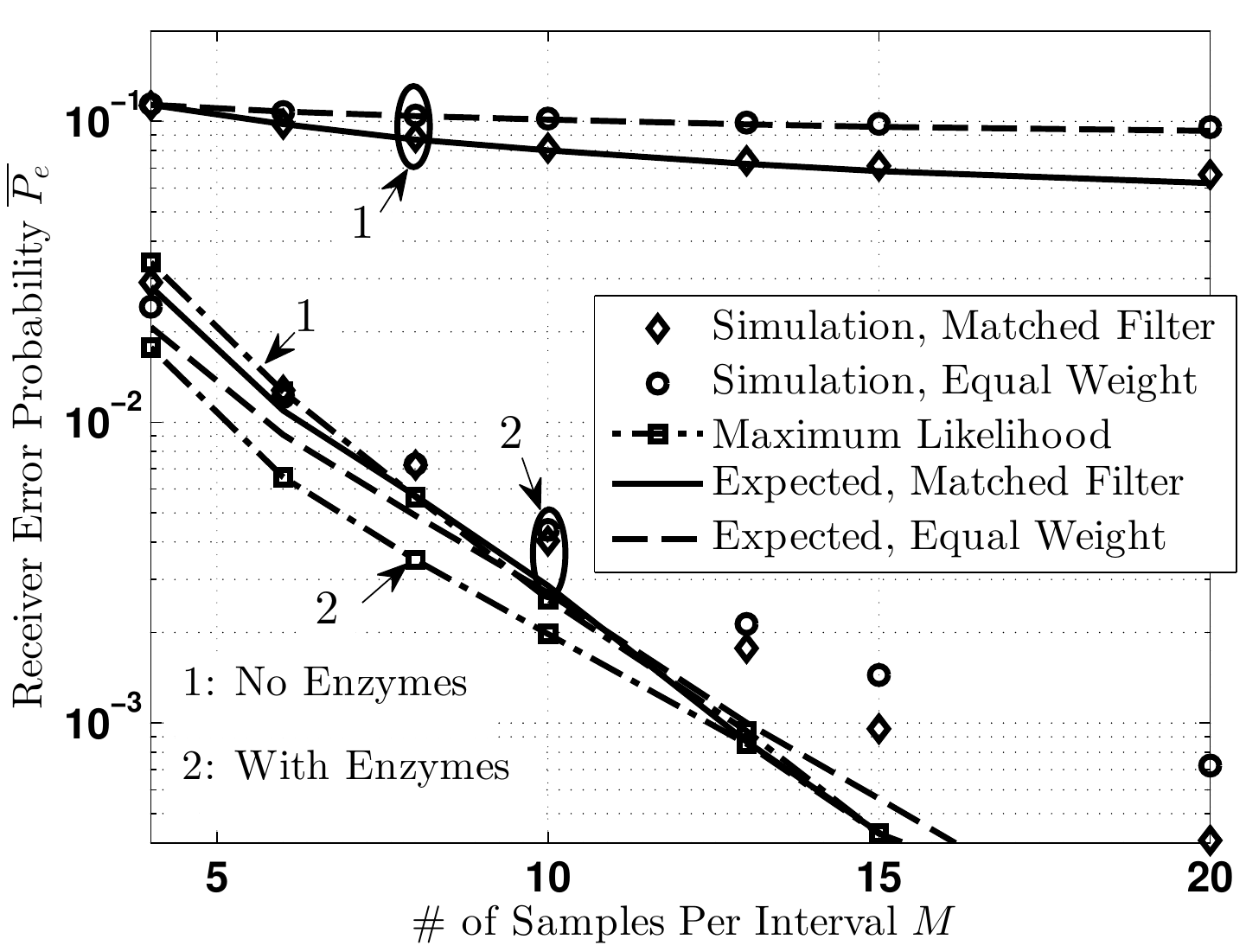}
	\caption{Receiver error probability as a function of $\M$
	when \ISI\, is included, $\T = 100\,\mu\second$,
	and enzymes are added to mitigate the impact of \ISI.}
	\label{fig5}
	\end{figure}
}

\newcommand{\figSix}[1]{
	\begin{figure}[#1]
	\centering
	\includegraphics[width=\linewidth]
	{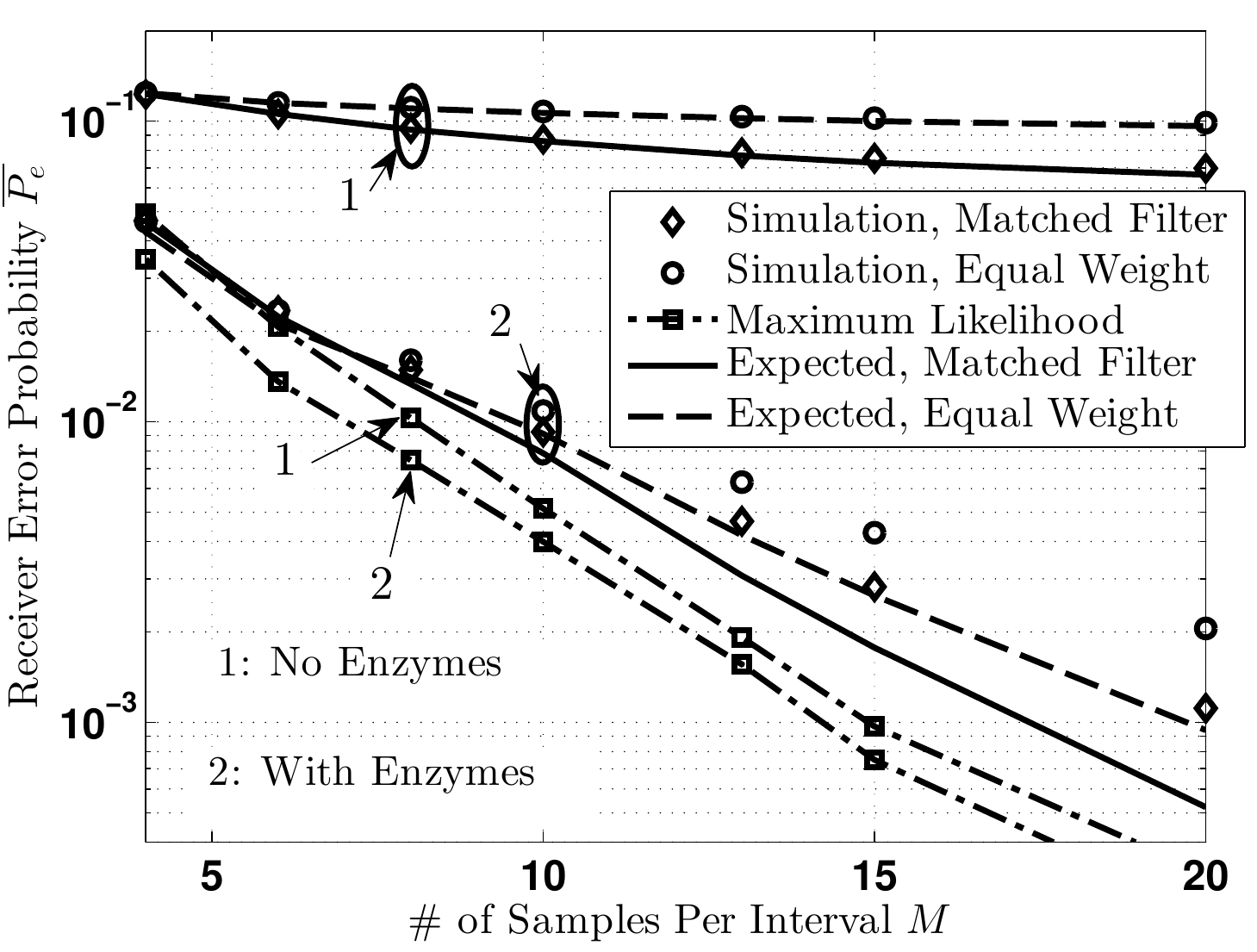}
	\caption{Receiver error probability as a function of $\M$
	when \ISI\, is included, $\T = 100\,\mu\second$,
	enzymes are added to mitigate the impact of \ISI, and an additive noise
	source is present ($\Nnoisetavg{t} = 1$ without enzymes and
	$\Nnoisetavg{t} = 0.5$ when enzymes are present).}
	\label{fig6}
	\end{figure}
}

\newcommand{\figSeven}[1]{
	\begin{figure}[#1]
	\centering
	\includegraphics[width=\linewidth]
	{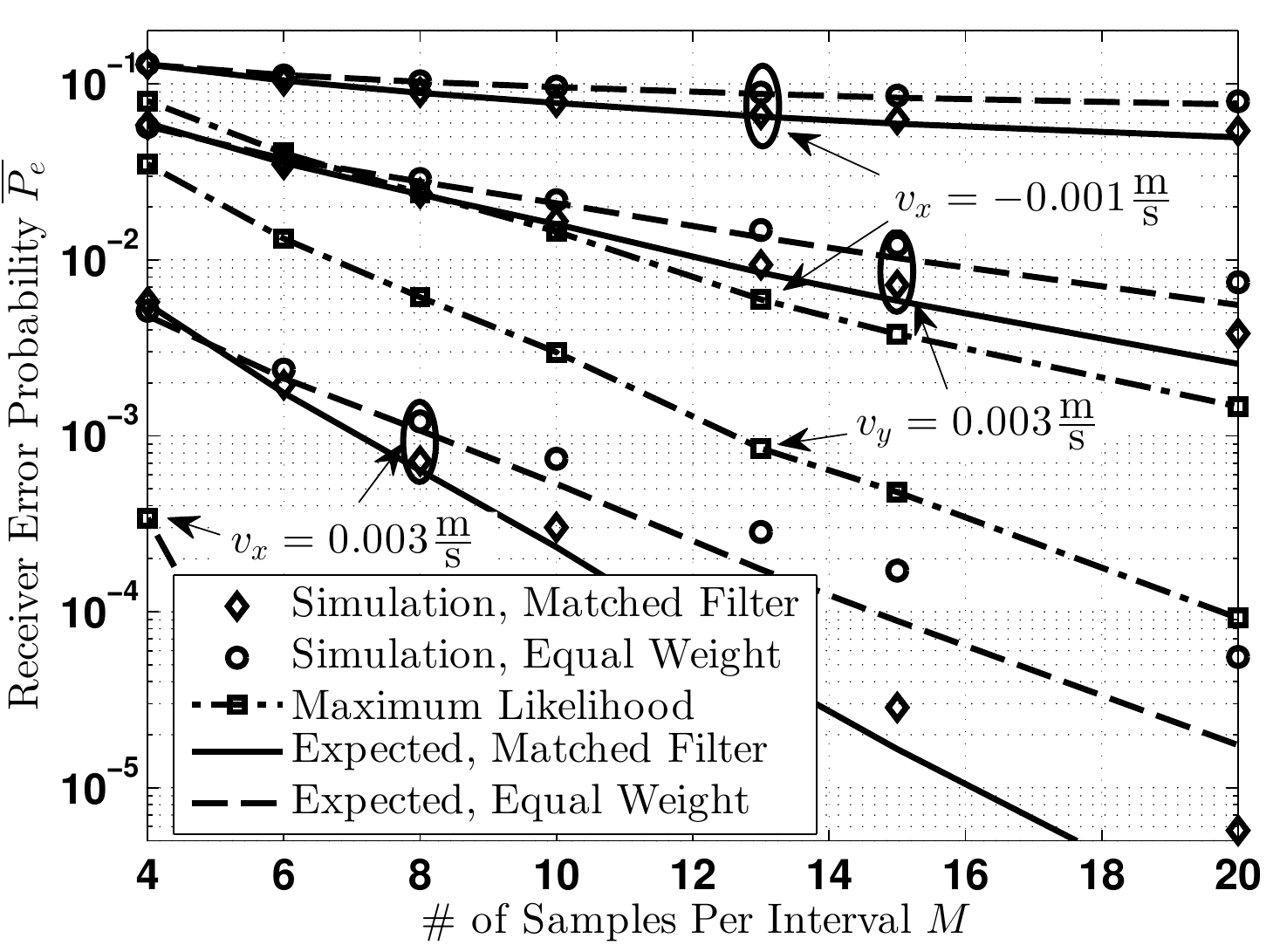}
	\caption{Receiver error probability as a function of $\M$
	when \ISI\, is included, $\T = 100\,\mu\second$,
	$\Nnoisetavg{t} = 1$, and different degrees of flow are present.}
	\label{fig7}
	\end{figure}
}

\maketitle

\begin{abstract}
In this paper, we perform receiver design for a diffusive
molecular communication environment. Our model includes flow
in any direction, sources of information molecules
in addition to the transmitter, and
enzymes in the propagation environment to mitigate
intersymbol interference. We characterize the mutual information
between receiver observations to show how often independent
observations can be made. We derive the maximum likelihood
sequence detector to provide a lower bound on the
bit error probability. We propose the family of weighted
sum detectors for more practical implementation and derive
their expected bit error probability. Under certain conditions,
the performance of the optimal
weighted sum detector is shown to be equivalent to a matched filter.
Receiver simulation results show the tradeoff in detector complexity
versus achievable bit error probability,
and that a slow flow in any direction can improve the
performance of a weighted sum detector.
\end{abstract}


\section{Introduction}

\PARstart{M}{olecular} communication is a promising design strategy
for the transfer of information between devices that have functional
components on the order of nanometers in size. The ability for
such devices to communicate would significantly expand their cumulative
potential, thereby enabling applications such as cooperative diagnostics
and drug delivery in biomedicine, bottom-up component design in
manufacturing, and sensitive environmental monitoring; see
\cite{RefWorks:540, RefWorks:608}. Molecular communication uses
physical molecules as information carriers, which are released by
the transmitter and propagate to an intended receiver. Propagation methods
include active schemes, where energy is consumed to direct the information
molecules towards the receiver, and passive schemes, where the transmitter
has no direct influence on the path taken by the released molecules.
Passive schemes, such as free diffusion, are simpler to implement and
more easily enable the formation of \emph{ad hoc} networks between devices.

Diffusion-based communication is commonly found in cellular systems
when small molecules need to quickly travel short distances; see
\cite[Ch. 4]{RefWorks:588}. Due to its simplicity and its current
implementation in nature, diffusion has also
often been chosen by communication researchers for the design of
synthetic nanonetworks. The diffusive channel has a number
of characteristics that distinguish it from conventional communication
channels. For example, it features randomly-moving molecules instead
of propagating electromagnetic radiation, so the propagation time
critically depends
on the distance between the transmitter and receiver
and there can be significant intersymbol interference (\ISI) due to
the lingering presence of emitted molecules.

Most existing literature that has studied the potential for transmitting
digital information over a diffusive channel has focused on numerically
characterizing the capacity in bits per second or bits per use, cf. e.g.
\cite{RefWorks:512, RefWorks:534,RefWorks:558,RefWorks:607,RefWorks:652}.
Recently, an upper-bound on capacity was analytically derived in \cite{RefWorks:687}.
Various strategies from conventional communications have
also been proposed to improve communication via diffusion, including
channel coding in \cite{RefWorks:668},
emitting Gaussian pulses in \cite{RefWorks:677}, and network coding
in \cite{RefWorks:695}.
However, the feasibility of such strategies is unknown.
We anticipate that
individual transceivers would have limited computational abilities, therefore
it is of interest to consider simple modulation schemes and then assess how
simple detectors perform in comparison to optimally designed detectors. Such
analysis would provide valuable insight into the practical design of diffusive
communication networks.

In this paper, we consider optimal and suboptimal receiver detection schemes
for a 3-dimensional stochastic diffusion environment between a single transmitter and
receiver. Our physical model includes steady uniform flow in any arbitrary direction,
as flow is likely to be present in any environment where molecular communication
is deployed; see \cite{RefWorks:540}.
We also consider
noise sources that release information molecules in addition to those released
by the transmitter. Generally, these sources could be interference from
similar communication links, or the result of
other chemical processes that produce information molecules as their output. We perform a
precise characterization of such noise sources in \cite{RefWorks:764}; here, we only
assume that their cumulative impact on the receiver can be characterized as a Poisson
random variable with time-varying mean (which is sufficient to model
either or both of the example cases above).
We consider the presence of enzyme
molecules in the propagation environment that are capable of degrading the information
molecules, which we introduced in \cite{RefWorks:631, RefWorks:662} as a strategy
to mitigate \ISI\, without requiring additional computational
complexity at the receiver. The use of enzymes to reduce \ISI\, was also recently
proposed in \cite{RefWorks:711}.

The primary contributions of this paper are as follows:
\begin{enumerate}
    \item We derive the mutual information between consecutive observations
    made by the receiver in order to measure the independence
    of observations. The assumption of independent observations is
    in general not satisfied, particularly as the time between observations tends
    to zero. However, we require sample independence for tractability in the design
    of the optimal detector, so it
    is essential to develop criteria to measure independence.
    \item We derive the optimal sequence detector, in a maximum likelihood
    sense, to give a lower bound on the bit error probability for any
    detector at the receiver. In practice, we do not expect such a detector
    to be physically realizable, due to the memory and computational complexity required,
    even if the implementation is simplified with the Viterbi algorithm
    as described in \cite{RefWorks:175}.
    \item We introduce weighted sum detectors as suboptimal but more physically realizable.
	In fact, they are used in neurons, which
	sum inputs from synapses with weights based on the characteristics
	of each synapse and fire when the sum exceeds
	a certain threshold; see \cite[Ch. 12]{RefWorks:587}.
	We characterize weighted sum detectors by the number of samples taken by the receiver in each bit interval and the weights assigned to each observation.
	We derive the expected bit error rate of any weighted sum detector.
	We show that a
	particular selection of sample weights is equivalent to a matched filter, and compare
	this detector with the optimal single bit detector.	We also consider equal sample weights
	as a simpler weight	selection scheme that is analogous to energy detection in conventional communication. We note that the receiver
	in our previous work in \cite{RefWorks:662} can be interpreted as the simplest special case of a weighted sum detector, where an immediate decision is made after only one observation in a given bit interval.
\end{enumerate}

Related work concerning the optimal sequence and single bit detectors was reported
in \cite{RefWorks:644}, where the authors modeled a 2-dimensional environment without
flow or an external noise source. The received signal was analyzed as a Gaussian random
variable with time-varying mean, where the transmitter symbols are represented by different
molecules. This paper considers modulation with a single type of molecule
but with a more general physical
model, including the possibility of using enzymes in the propagation environment. We perform
the majority of our analysis assuming that the
received signal is a Poisson random variable with time-varying mean, which is more
easily satisfied and more accurate than the Gaussian approximation. Existing papers that
model flow provide detailed analysis but assume that flow is only possible in the
direction along a line joining the
transmitter and receiver, and furthermore that the physical model itself is one-dimensional,
cf. e.g. \cite{RefWorks:607, RefWorks:615}. This paper does not focus on the impact of flow,
but (with the recent exception of \cite{RefWorks:734}) our inclusion of flow in any arbitrary direction has not yet been considered in
the literature. 

It is important to acknowledge the major simplifying assumptions that are made for
our analysis. We do this to highlight the applicability of our work, and also to identify
areas for future study. These assumptions are as follows:
\begin{enumerate}
    \item We assume that the transmitter emits enough molecules for the
	observations made at the receiver to have a Poisson distribution\footnote{The number of molecules observed can be most accurately described by the binomial distribution, as we previously discussed in \cite{RefWorks:662}, and the Poisson approximation of the binomial distribution is commonly applied when the number of trials is large and the probability of success is small; see \cite{RefWorks:725}. Here, each released molecule represents one trial and the probability of success is that molecule's probability of being observed at the receiver at some time.}.
	\item We assume that the transmitter emits insufficient molecules 
	to change the diffusion coefficient
	of the propagation medium. Strictly speaking, the diffusion coefficient is a function of the local concentration of information molecules (see \cite{RefWorks:742}), so here the coefficient should vary both spatially and temporally. However, we assume that the information molecules are sufficiently dilute to ignore collisions between them in the propagation environment\footnote{In our simulations, the highest information molecule concentration is at the transmitter when the molecules are released. This concentration is $1.67\,\textnormal{M}$, which is still much lower than the solvent (water) molecule concentration, which is about $55\,\textnormal{M}$.}.
	\item The modulation scheme adopted by the transmitter is simple ON/OFF
	keying with a constant bit interval, where an impulse of molecules is
	released at the start of an interval to transmit a $1$ and the transmitter is
	``silent'' to transmit a $0$. We note that our results can be easily
	extended to pulse amplitude modulation, where the number of molecules
	released corresponds to a particular information symbol.	
	\item We assume	an ideal receiver that does not interact with the
	information molecules but is able to perfectly count the number of those
	molecules that are within its observation space, as
	in \cite{RefWorks:687}. We also assume that the receiver
	is synchronized in time with the transmitter, for example via an external signal
	as discussed in \cite{RefWorks:614}. We do so in order to focus
	on the effects of the propagation environment and for tractability.
	Recent work in
	\cite{RefWorks:717,RefWorks:723} has incorporated the impact of a chemical
	reaction mechanism at the receiver where the observable molecules are the output
	of a reaction involving the molecules emitted by the transmitter. However,
	whether such a model can lead to a closed-form time domain expression for
	the signal observed at the receiver remains an open problem.
	\item For the purposes of deriving the mutual information of consecutive receiver
	observations, we assume that the receiver is spherical (or, equivalently,
	hemispherical if it is mounted to an infinite plane that forms an elastic boundary
	of a semi-infinite environment). We focus on the sphere because
	this shape is naturally formed by membrane bilayers; see
	\cite[Ch. 11]{RefWorks:588}. These bilayers can have molecule receptors embedded
	in them as part of a functioning nanomachine, such as in a cell.
\end{enumerate}

The rest of this paper is organized as follows. The details of the
transmission environment are described in Section~\ref{sec_model}.
In Section~\ref{sec_obs}, we present the channel impulse response
and describe the time-varying signal that is observed at the receiver.
In Section~\ref{sec_indp}, we derive the mutual information between
consecutive observations made by the receiver. We derive the
optimal sequence detector in a maximum likelihood sense in
Section~\ref{sec_opt}, assuming
independent observations.
In Section~\ref{sec_subopt}, we introduce weighted sum
detectors, which are suboptimal for receiving a sequence
of bits, but which may be more easily realizable for bio-inspired
nanoscale and microscale communication networks. Numerical results
showing detector performance are described in Section~\ref{sec_num},
and conclusions are drawn in Section~\ref{sec_concl}.

\section{System Model}
\label{sec_model}

There is a transmitter fixed at the origin of an infinite,
3-dimensional fluid environment of uniform constant temperature
and viscosity. The receiver is a sphere with radius $\robs$ and
volume $\Vobs$. It is centered at location $\{\x_0,0,0\}$
(without loss of generality) where
$\rad{0}$ is the vector from the origin to $\{\x_0,0,0\}$.
The receiver is a passive observer
that does not impede diffusion or initiate chemical reactions.
We also note that, by symmetry, the concentrations observed in
this environment are equivalent (by a factor of 2)
to those in the semi-infinite
case where $\z \ge 0$ is the fluid environment, the $\x\y$-plane is an elastic
boundary, and the receiver is a hemisphere whose circle face lies on the boundary;
see \cite[Eq. (2.7)]{RefWorks:586}.
This equivalent environment could describe, for example, a small transmitter
and receiver mounted along the wall of a blood large vessel.
A steady uniform flow (or drift) exists in an
arbitrary direction defined by its velocity component along each dimension, i.e.,
$\{\vx{\x},\vx{\y},\vx{\z}\}$. Analytically, this is the simplest type of flow. Other flows of interest but outside the scope of this work include laminar flow, where successive layers of fluid slide over one another without mixing, as is common in environments such as small blood vessels; see \cite[Ch. 5]{RefWorks:587}.

The transmitter is a source of information molecules called $\A$ molecules. They
have a negligible natural degradation rate into $\AP$ molecules, but are able to
degrade much more quickly if they bind to enzyme molecules called $\En$ 
molecules. We assume that $\A$ and $\En$ molecules react via the Michaelis-Menten
reaction mechanism, described as
\begin{align}
\label{k1_mechanism}
\En + \A &\xrightarrow{\kth{1}} \EA, \\
\label{kminus1_mechanism}
\EA &\xrightarrow{\kth{-1}} \En + \A, \\
\label{k2_mechanism}
\EA &\xrightarrow{\kth{2}} \En + \AP,
\end{align}
where $\EA$ is the intermediate formed by the binding of an $\A$ molecule to an
$\En$ molecule, and $\kth{1}$,
$\kth{-1}$, and $\kth{2}$ are the reaction rates for the reactions as shown
in (\ref{k1_mechanism})-(\ref{k2_mechanism})
with units $\molecule^{-1}\metre^3\,\second^{-1}$, $\second^{-1}$, and
$\second^{-1}$, respectively. This mechanism is generally accepted as the fundamental
mechanism for enzymatic reactions, though it is not the only mechanism possible;
see \cite[Ch. 10]{RefWorks:585}.

In addition to the transmitter, we assume that the environment has other sources
of $\A$ molecules, either via interference from other communication links or via
some other chemical process that generates $\A$ molecules. We only assume that
we know the cumulative impact of all noise sources on the received signal, and
that we can characterize this impact as a Poisson random variable with time-varying
mean. In our simulations,
we will limit ourselves to an additive noise variable with constant mean (we perform a more precise characterization of noise and interference in \cite{RefWorks:764}).
We emphasize that this additive noise source is distinct from the randomness in the
number of $\A$ molecules observed by the receiver due to diffusion.

We use a common notation to describe the different molecule types in our model.
The number of molecules of species $\X$ is given by $\Nx{\X}$, and its
concentration at the point defined by vector $\rad{}$ and at time $t$ in
$\molecule\cdot\metre^{-3}$ is $\CxFun{\X}{\rad{}}{t}$ (generally written as
$\Cx{\X}$ for compactness). We assume that every molecule diffuses independently
of every other molecule, and that every molecule is spherical, so that they all
diffuse with constant diffusion coefficient $\Dx{\X}$,
found using the Einstein relation as \cite[Eq. (4.16)]{RefWorks:587}
\begin{equation}
\label{JUN12_60}
\Dx{\X} = \frac{\bolt\temp}{6\pi \visc \Ri{\X}},
\end{equation}
where $\bolt$ is the Boltzmann constant ($\bolt = 1.38 \times 10^{-23}$ J/K),
$\temp$ is the temperature in kelvin, $\visc$ is the viscosity of the
propagation environment, and $\Ri{\X}$ is the
molecule radius.
We note that the accuracy of the Einstein relation is limited in practice (see \cite[Ch. 5]{RefWorks:742}), and we only use it here to get a sense of appropriate values for $\Dx{\X}$; typically, $\Dx{\X}$ is found via experiment.
The diffusion (i.e., displacement) of a single molecule along
one dimension is a Gaussian random variable with variance $2\Dx{\X} t$, where
$t$ is the diffusing time in seconds; see \cite[Eq. (4.6)]{RefWorks:587}. The presence of steady uniform flow means that every diffusing molecule has a net \emph{constant} displacement due to the flow in addition to the \emph{random} displacement due to diffusion.

The transmitter has a sequence of binary data $\dataSeq = \{\data{1},\data{2},\dots\}$ to
send to the receiver, where $\data{j}$ is the $j$th information bit and
$\Pr(\data{j} = 1) = \Pone$. The transmitter
uses binary modulation and transmission intervals of duration $\T$ seconds. To send
a binary $1$, $\Nemit$ $\A$ molecules are released at the start of the bit interval.
To send a binary $0$, no molecules are released. This method is effectively ON/OFF
keying, but we place all signal energy at the start of the bit interval in order to
mitigate \ISI. There is a constant total enzyme concentration $\Cx{\Etot}$ everywhere.
We facilitate simulations by considering $\Ne$ $\En$ molecules
that are initially uniformly
distributed throughout and bounded within a finite volume $\Ve$ that is sufficiently large
to assume that it is infinite in size (although we do not restrict the motion of $\A$
molecules to within $\Ve$, and $\EA$ molecules probabilistically undergo reaction
(\ref{kminus1_mechanism}) or (\ref{k2_mechanism}) if they hit the boundary).

\section{Receiver Signal}
\label{sec_obs}

The receiver is an ideal passive observer that is synchronized with the transmitter.
The signal that it ``observes'' is a discrete counting signal that represents the
number of molecules present within $\Vobs$ at the time when the observation is made.
This signal, $\Nobst{t}$, has two components: $\Ntxt{t}$, or the number of observed
molecules that originated at the transmitter, and $\Nnoiset{t}$, or the number
of observed molecules that originated at the additive noise sources.
We emphasize that even without additive noise sources,
i.e., $\Nnoisetavg{t} = 0$, $\Nobst{t}$ will still be noisy
due to the random motion of the molecules emitted by the transmitter.
In this section, we first describe $\Nobst{t}$ when there is only a single emission by the
transmitter in the first bit interval, i.e., the channel impulse response scaled by
$\Nemit$. Next, we characterize $\Nobst{t}$ as a random variable due to transmitter
sequence $\dataSeq$.
The signal model
is a generalization of the receiver signal presented
in \cite{RefWorks:662}; here, we add the impacts of flow and external noise sources.
We have already assumed that we characterize $\Nnoiset{t}$ as
a Poisson random variable with time-varying mean, and that the diffusion of molecules is
independent. Thus, in the following we only need to characterize $\Ntxt{t}$.

\subsection{Channel Impulse Response}

The general reaction-diffusion equation for species $\X$ is
\cite[Eq. (8.12.1)]{RefWorks:602}
\begin{equation}
\label{JUN12_33}
\pbypx{\Cx{\X}}{t} = \Dx{\X}\nabla^2\Cx{\X} + \ux{\Cx{\X},\rad{},t},
\end{equation}
where $\ux{\cdot}$ is the reaction term. Using the principles of
chemical kinetics as described in \cite[Ch. 9]{RefWorks:585}, the system
of partial differential equations for our environment \emph{without flow} is
\cite[Eqs. (6)-(8)]{RefWorks:662}
\begin{align}
\label{AUG12_39}
\pbypx{\Cx{\A}}{t} = &\; \Da\nabla^2\Cx{\A} -\kth{1}\Cx{\A}\Cx{\En} +
\kth{-1}\Cx{\EA}, \\
\label{AUG12_40}
\pbypx{\Cx{\En}}{t} = &\; \De\nabla^2\Cx{\En} -\kth{1}\Cx{\A}\Cx{\En} +
\kth{-1}\Cx{\EA} + \kth{2}\Cx{\EA},\\
\label{AUG12_41}
\pbypx{\Cx{\EA}}{t} = &\; \Di\nabla^2\Cx{\EA} +\kth{1}\Cx{\A}\Cx{\En} -
\kth{-1}\Cx{\EA} - \kth{2}\Cx{\EA}.
\end{align}

To incorporate steady uniform flow,
we only need to change the diffusion coefficient terms, as
described in \cite[Ch. 4]{RefWorks:630}. Specifically, for the $\A$ molecules we replace
$\Da\nabla^2\Cx{\A}$ with
\begin{equation}
\Da\nabla^2\Cx{\A} - \vx{\x}\pbypx{\Cx{\A}}{\x} - \vx{\y}\pbypx{\Cx{\A}}{\y}
 - \vx{\z}\pbypx{\Cx{\A}}{\z},
\label{EQ13-04-21}
\end{equation}
and make analogous substitutions for $\En$ and $\EA$. The system of
equations above has no closed-form analytical solution, even under our
boundary conditions of
an impulsive point source and unbounded environment. In \cite[Eq. (10)]{RefWorks:662},
we derived a lower bound on the \emph{expected} point concentration of
$\A$ molecules at a distance $\radmag{}$ from an impulsive point source
that emits at time $t = 0$. This lower bound is
\begin{equation}
\label{JUN12_47}
\Cx{\A} \ge \frac{\Nemit}{(4\pi \Da
t)^{3/2}}\EXP{-\kth{}\Cx{\Etot}t - \frac{\radmag{}^2}{4\Da t}},
\end{equation}
where
\begin{equation}
\kth{} = \left\{
 \begin{array}{rl}
  \kth{1} & \text{for a strict lower bound},\\
  \frac{\kth{1}\kth{2}}{\kth{-1} + \kth{2}} & \text{for an approximation},
 \end{array} \right.
\end{equation}
and the approximation for $\kth{}$ is much more accurate than the lower bound if
$\kth{-1} + \kth{2} \approx \kth{2}$ is not satisfied. The lower bound becomes tight as
$\kth{2} \to\infty$ and $\kth{-1}\to 0$, and it is also \emph{equivalent} to the expected concentration
if $\A$ molecules degrade \emph{directly} into $\AP$ molecules at rate $\kth{}\Cx{\Etot}$.
To include the effect of flow, we simply
set $\radmag{}^2 = (\x - \vx{x}t)^2 + (\y - \vx{y}t)^2 + (\z - \vx{z}t)^2$ for the
concentration at point $\{\x,\y,\z\}$.

For clarity of exposition and for tractability in the presence of flow
in any direction,
we apply the uniform concentration assumption
to calculate $\Ntxt{t}$ from $\Cx{\A}$,
where we assume that the expected concentration
throughout the receiver is
equal to that expected at the center of the receiver.
We showed in \cite{RefWorks:706}
that the validity of this assumption in the no-flow case improves as the receiver is placed further
from the transmitter.
We also assume that (\ref{JUN12_47}) is satisfied with equality (which is
always true when there are no enzymes present) and thus find that
the \emph{expected} channel impulse response at the receiver due to the transmitter,
$\Ntxtavg{t}$, is
\ifOneCol
\begin{equation}
\Ntxtavg{t} = \Vobs\CxFun{\A}{(\x_0 - \vx{x}t)^2 + (\vx{y}t)^2 + (\vx{z}t)^2}{t}
= \Vobs\CxFun{\A}{\radmag{eff}}{t},
\label{JUN12_47_uca}
\end{equation}
\else
\begin{align}
\Ntxtavg{t} = &\; \Vobs\CxFun{\A}{\sqrt{(\x_0 - \vx{x}t)^2 + (\vx{y}t)^2 + (\vx{z}t)^2}}{t}
\nonumber \\
= &\; \Vobs\CxFun{\A}{\radmag{eff}}{t},
\label{JUN12_47_uca}
\end{align}
\fi
where $\radmag{eff} = \sqrt{(\x_0 - \vx{x}t)^2 + (\vx{y}t)^2 + (\vx{z}t)^2}$
is the \emph{effective} distance
between the transmitter and the center of the receiver, considering the flow.

For a single molecule emitted by the transmitter at time $t = 0$, the probability that it is
within $\Vobs$ at time $t$, $\Pobsx{t}$, is given by (\ref{JUN12_47_uca}) where
$\Nemit = 1$, i.e.,
\begin{equation}
\Pobsx{t} = \frac{\Vobs}{(4\pi \Da
t)^{3/2}}\EXP{-\kth{}\Cx{\Etot}t - \frac{\radmag{eff}^2}{4\Da t}}.
\label{JUN12_47_pobs}
\end{equation}

We showed in \cite{RefWorks:662} that $\Ntxt{t}$ due to one emission
of $\Nemit$ molecules at the transmitter
can be accurately approximated as a Poisson random variable with time-varying mean given
by $\Ntxtavg{t}$ in (\ref{JUN12_47_uca}). Since $\Nnoiset{t}$ is also
a Poisson random variable with time-varying mean, then $\Nobst{t}$ is the sum of two
Poisson random variables. The sum of independent Poisson random variables is also a Poisson random
variable whose mean is the sum of the means of the individual variables; see
\cite[Ch. 5.2]{RefWorks:725}. Thus, the signal observed is a Poisson
random variable with mean $\Nnoisetavg{t} + \Ntxtavg{t}$ with
$\Ntxtavg{t}$ as given by (\ref{JUN12_47_uca}).

\subsection{Complete Receiver Signal}

In general, the receiver can observe molecules that were emitted at the start of the
current or any prior bit interval, depending on $\dataSeq$. We assume that the
behavior of individual $\A$ molecules is independent, so we can immediately write
\begin{equation}
\label{EQ13_04_02notavg}
\Ntxt{t} = \sum_{j=1}^{\floor{\frac{t}{\T}+1}}\Ntxt{t;j},
\end{equation}
where $\Ntxt{t;j}$ is the number of molecules observed at time $t$ that were emitted
at the start of the $j$th bit interval. Of course, $\Ntxt{t;j} = 0\; \forall\; t$ if
$\data{j} = 0$. $\Ntxt{t}$ is still a sum of Poisson random variables, and so
$\Nobst{t}$ is a Poisson random variable with mean
\begin{equation}
\Nobsavgt = \Ntxtavg{t} + \Nnoisetavg{t},
\label{EQ13_05_28}
\end{equation}
where
\begin{equation}
\label{EQ13_04_02}
\Ntxtavg{t} = \Nemit\sum_{j=1}^{\floor{\frac{t}{\T}+1}}\data{j}\Pobsx{t-(j-1)\T}.
\end{equation}

Thus, the probability density function (\PDF) of $\Nobst{t}$ is
\begin{equation}
\Pr\left(\Nobst{t} = \thresh \right) =
\frac{\Nobsavg{t}^\thresh\EXP{-\Nobsavg{t}}}{\thresh!},
\label{EQ13_04_03}
\end{equation}
and the cumulative density function (\CDF) is
\begin{equation}
\Pr\left(\Nobst{t} < \thresh \right) =
\EXP{-\Nobsavg{t}}
\sum_{i=0}^{\thresh-1}
\frac{\Nobsavg{t}^i}{i!}.
\label{EQ13_04_04}
\end{equation}

\section{Independence of Receiver Observations}
\label{sec_indp}

In this section, we study the assumption that
all observations made at the receiver are independent of each
other. Intuitively, this will not be true if the time between samples
tends to zero; if the time between samples is infinitesimally small,
then there is insufficient time for either the molecules observed
in one sample to leave the receiver or for any new molecules to arrive.
We will measure independence using mutual information, which is a measure
of how the knowledge of one variable influences the prediction of a
second variable; see \cite{RefWorks:707}.
Specifically, for two discrete random variables $X$ and $Y$
taking values $x$ and $y$,
respectively, the mutual information $\Ix{}{X;Y}$ is
\cite[Eq. (2.28)]{RefWorks:707}
\ifOneCol
\begin{equation}
\Ix{}{X;Y} = \sum\limits_x\sum\limits_y \Pr\left(X=x,Y=y\right)
\log\frac{\Pr\left(X=x,Y=y\right)}{\Pr(X=x)\Pr(Y=y)},
\label{EQ13_04_18_gen}
\end{equation}
\else
\begin{align}
\Ix{}{X;Y} = & \sum\limits_x\sum\limits_y \Pr\left(X=x,Y=y\right)
\nonumber \\
& \times\log\frac{\Pr\left(X=x,Y=y\right)}{\Pr(X=x)\Pr(Y=y)},
\label{EQ13_04_18_gen}
\end{align}
\fi
where the summations are over all possible values of $x$ and $y$. If
the variables $X$ and $Y$ are independent, then $\Ix{}{X;Y} = 0$,
i.e., knowing $X$ does not provide any greater certainty in predicting
the value of $Y$, and vice versa. If $X$ and $Y$ are measures of the
same variable over time, and the time between measures
is infinitesimally small, then knowing $X$ lets us predict $Y$ perfectly and
mutual information is maximized.

We focus on the independence of two consecutive receiver observations,
since these are the most likely to be dependent.
We define these observations
to be at times $t_1$ and $t_2$, such that $t_2 > t_1$. 
For clarity, we consider only molecules
that were emitted by the transmitter at time $t = 0$, so the unconditional
time-varying \PDF\, describing whether a molecule is within $\Vobs$ is given
by $\Pobsx{t}$ in (\ref{JUN12_47_pobs}). Of course, if no molecules are emitted and there are no noise sources, then the mutual information
between samples of molecules from the transmitter would be zero because
every sample would be $0$.
For tractability, we also assume no
flow in this section, though we expect that the presence of flow
can only decrease the
mutual information of the two observations.

Let the observations at times $t_1$ and $t_2$ be $\sx{1}$ and $\sx{2}$,
respectively. From probability theory, the joint probability distribution
of $\Nobst{t_1}$ and $\Nobst{t_2}$ can be evaluated from the conditional
probability distribution, i.e.,
\ifOneCol
\begin{equation}
\Pr\left(\Nobst{t_1} = \sx{1}, \Nobst{t_2} = \sx{2}\right) =
\Pr\left(\Nobst{t_2} = \sx{2} \mid \Nobst{t_1} = \sx{1}\right)
\Pr\left(\Nobst{t_1} = \sx{1}\right).
\label{EQ13_05_15}
\end{equation}
\else
\begin{align}
& \Pr\left(\Nobst{t_1} = \sx{1}, \Nobst{t_2} = \sx{2}\right) = \nonumber \\
& \Pr\left(\Nobst{t_2} = \sx{2} \mid \Nobst{t_1} = \sx{1}\right)
\Pr\left(\Nobst{t_1} = \sx{1}\right).
\label{EQ13_05_15}
\end{align}
\fi

The mutual information is then
\ifOneCol
\begin{align}
&\Ix{}{\Nobst{t_1};\Nobst{t_2}} =
\sum\limits_{\sx{1}}\sum\limits_{\sx{2}}
\Pr\left(\Nobst{t_1}=\sx{1},\Nobst{t_2}=\sx{2}\right) \nonumber \\
& \qquad\times\log\frac{\Pr\left(\Nobst{t_1}=\sx{1},\Nobst{t_2}=\sx{2}\right)}
{\Pr(\Nobst{t_1}=\sx{1})\Pr(\Nobst{t_2}=\sx{2})},
\label{EQ13_04_18}
\end{align}
\else
\begin{align}
&\Ix{}{\Nobst{t_1};\Nobst{t_2}} = \nonumber \\
& \qquad\sum\limits_{\sx{1}}\sum\limits_{\sx{2}}
\Pr\left(\Nobst{t_1}=\sx{1},\Nobst{t_2}=\sx{2}\right) \nonumber \\
& \qquad\times\log\frac{\Pr\left(\Nobst{t_1}=\sx{1},\Nobst{t_2}=\sx{2}\right)}
{\Pr(\Nobst{t_1}=\sx{1})\Pr(\Nobst{t_2}=\sx{2})},
\label{EQ13_04_18}    
\end{align}
\fi
where the range of values for the observations $\sx{1}$ and $\sx{2}$ in the summations should account for all observations that have a non-negligible probability of occurring.
The marginal probabilities (i.e., of a single variable)
in (\ref{EQ13_04_18}) can be evaluated using (\ref{EQ13_04_03}) and used to determine appropriate ranges for $\sx{1}$ and $\sx{2}$.
The remaining step to evaluate (\ref{EQ13_04_18}) is to derive the
conditional probability in (\ref{EQ13_05_15}). 
We can derive this probability if we have an expression for $\pstay{\deltObs}$,
the probability that an information molecule that is observed at the
receiver at $t_1$ is later at the receiver at $\deltObs = t_2 - t_1$.
We assume that an information molecule observed at $t_1$ is randomly located
within the receiver sphere, following the uniform concentration assumption.
Then, from \cite[Eq. (3.8)]{RefWorks:586}, the point
concentration $\CxFun{\A}{r}{\deltObs}$
due to this single molecule at a distance $r$ from
the center of the receiver, without enzymes present, is
\ifOneCol
\begin{align}
\CxFun{\A}{r}{\deltObs} = &\; \frac{3}{8\pi\robs^3}\left[
\ERF{\frac{\robs-r}{2\sqrt{\Dx{\A}\deltObs}}} +
\ERF{\frac{\robs+r}{2\sqrt{\Dx{\A}\deltObs}}}\right] \nonumber
\\
& +
\frac{3}{4\pi\robs^3 r}\sqrt{\frac{\Dx{\A}\deltObs}{\pi}}
\bigg[\EXP{-\frac{(\robs+r)^2}{4\Dx{\A}\deltObs}}
- \EXP{-\frac{(\robs-r)^2}{4\Dx{\A}\deltObs}}\bigg],
\label{MAR13_01}
\end{align}
\else
\begin{align}
\CxFun{\A}{r}{\deltObs} = &\; \frac{3}{8\pi\robs^3}\left[
\ERF{\frac{\robs-r}{2\sqrt{\Dx{\A}\deltObs}}} +
\ERF{\frac{\robs+r}{2\sqrt{\Dx{\A}\deltObs}}}\right] \nonumber
\\
& +
\frac{3}{4\pi\robs^3 r}\sqrt{\frac{\Dx{\A}\deltObs}{\pi}}
\bigg[\EXP{-\frac{(\robs+r)^2}{4\Dx{\A}\deltObs}} \nonumber \\
& - \EXP{-\frac{(\robs-r)^2}{4\Dx{\A}\deltObs}}\bigg],
\label{MAR13_01}
\end{align}
\fi
where the error function is \cite[Eq. (8.250.1)]{RefWorks:402}
\begin{equation}
\label{APR12_32}
\ERF{x} = \frac{2}{\pi}\int_0^x \EXP{-y^2} dy,
\end{equation}
and we can easily account for enzymes by multiplying (\ref{MAR13_01})
by a factor of $\EXP{-\kth{}\Cx{\Etot}\deltObs}$.
What follows in the remainder of this section
is valid with or without this factor included.

To derive $\pstay{\deltObs}$, we must integrate $\CxFun{\A}{r}{\deltObs}$ over
the entire receiver volume, i.e.,
\ifOneCol
\begin{equation}
\pstay{\deltObs} = \int\limits_0^{\robs}
\int\limits_{0}^{2\pi}
\int\limits_{0}^{\pi}
\CxFun{\A}{r}{\deltObs}r^2\sin\theta
d\theta d\phi dr
 = 4\pi\int\limits_0^{\robs}
\CxFun{\A}{r}{\deltObs}r^2 dr.
\label{MAR13_03}
\end{equation}
\else
\begin{align}
\pstay{\deltObs} & = \int\limits_0^{\robs}
\int\limits_{0}^{2\pi}
\int\limits_{0}^{\pi}
\CxFun{\A}{r}{\deltObs}r^2\sin\theta
d\theta d\phi dr \nonumber \\
& = 4\pi\int\limits_0^{\robs}
\CxFun{\A}{r}{\deltObs}r^2 dr.
\label{MAR13_03}
\end{align}
\fi

We now present the following theorem:

\begin{theorem}[$\pstay{\deltObs}$ for one molecule]
\label{theorem_pstay}
The probability of an information molecule being inside the receiver
at time $\deltObs$ after it was last observed inside the receiver is given by
\ifOneCol
\begin{equation}
\pstay{\deltObs} = \ERF{\frac{\robs}{\sqrt{\Dx{\A}\deltObs}}} +
\frac{1}{\robs}\sqrt{\frac{\Dx{\A}\deltObs}{\pi}}
\bigg[\Big(1 - \frac{2\Dx{\A}\deltObs}{\robs^2}\Big)
\EXP{\frac{-\robs^2}{\Dx{\A}\deltObs}} + \frac{2\Dx{\A}\deltObs}{\robs^2} - 3\bigg],
\label{MAR13_31}
\end{equation}
\else
\begin{align}
\pstay{\deltObs} = & \ERF{\frac{\robs}{\sqrt{\Dx{\A}\deltObs}}} +
\frac{1}{\robs}\sqrt{\frac{\Dx{\A}\deltObs}{\pi}} \nonumber \\
& \times\!
\bigg[\Big(1 - \frac{2\Dx{\A}\deltObs}{\robs^2}\Big)
\EXP{\frac{-\robs^2}{\Dx{\A}\deltObs}}\! + \frac{2\Dx{\A}\deltObs}{\robs^2} - 3\bigg],
\label{MAR13_31}
\end{align}
\fi
where (\ref{MAR13_31}) is multiplied by $\EXP{-\kth{}\Cx{\Etot}\deltObs}$ if
enzymes are present.
\end{theorem}
\begin{IEEEproof}
Please refer to the Appendix.
\end{IEEEproof}

The probability that a molecule observed at $t_1$ is outside the observer at
$t_2$ is $\pleave{\deltObs} = \EXP{-\kth{}\Cx{\Etot}\deltObs} -\pstay{\deltObs}$.

To derive an expression for (\ref{EQ13_05_15}),
we also require an expression for
$\parrive{t_1,t_2}$, the unconditional probability that
an information molecule that is outside the receiver at $t_1$ is inside the
receiver at $t_2$. Intuitively, this is equal to the unconditional
probability of the molecule being inside the receiver at $t_2$, minus
the probability of the molecule being inside the receiver at $t_1$ and
still inside the receiver at $t_2$, i.e.,
\begin{equation}
\parrive{t_1,t_2} = \Pobsx{t_2} - \Pobsx{t_1}\pstay{\deltObs}.
\label{FEB13_18}
\end{equation}

The conditional probability in (\ref{EQ13_05_15}) must consider every possible
number of arrivals and departures of molecules from the receiver in order to
result in a net change of $\sx{2}-\sx{1}$ information molecules. In other words,
\ifOneCol
\begin{equation}
\Pr\left(\Nobst{t_2} = \sx{2} \mid \Nobst{t_1} = \sx{1}\right) =
\sum\limits_{i = \max(0,\sx{1}-\sx{2})}^{\sx{1}}
\Pr\left(i\, \textnormal{leave}\right)
\Pr\left(\sx{2}+i-\sx{1} \,\textnormal{arrive}\right),
\label{FEB13_16}
\end{equation}
\else
\begin{align}
&\Pr\left(\Nobst{t_2} = \sx{2} \mid \Nobst{t_1} = \sx{1}\right) = \nonumber \\
&\quad\sum\limits_{i = \max(0,\sx{1}-\sx{2})}^{\sx{1}}
\Pr\left(i \,\textnormal{leave}\right)
\Pr\left(\sx{2}+i-\sx{1} \,\textnormal{arrive}\right),
\label{FEB13_16}
\end{align}
\fi
where $\Pr\left(i \,\textnormal{leave}\right)$ is
conditioned on $\Nobst{t_1}$ as
\ifOneCol
\begin{equation}
\Pr\left(i \,\textnormal{leave} \mid \Nobst{t_1} = \sx{1} \right) =
\binom{\sx{1}}{i}\pleave{\deltObs}^i\left(1-\pleave{\deltObs}\right)^{\sx{1}-i},
\label{FEB13_17}
\end{equation}
\else
\begin{align}
&\Pr\left(i \,\textnormal{leave} \mid \Nobst{t_1} = \sx{1} \right) = \nonumber \\
&\quad\binom{\sx{1}}{i}\pleave{\deltObs}^i\left(1-\pleave{\deltObs}\right)^{\sx{1}-i},
\label{FEB13_17}
\end{align}
\fi
and $\Pr\left(i \,\textnormal{arrive}\right)$
is also conditioned on $\Nobst{t_1}$ as
\ifOneCol
\begin{equation}
\Pr\left(i \,\textnormal{arrive}\mid \Nobst{t_1} = \sx{1}\right) =
\binom{\Nx{\A}-\sx{1}}{i}\parrive{t_1,t_2}^i
\left[1-\parrive{t_1,t_2}\right]^{\Nx{\A}-\sx{1}-i}.
\label{FEB13_19}
\end{equation}
\else
\begin{align}
&\Pr\left(i \,\textnormal{arrive}\mid \Nobst{t_1} = \sx{1}\right) =
\nonumber \\
& \quad\binom{\Nx{\A}-\sx{1}}{i} \parrive{t_1,t_2}^i
\left[1-\parrive{t_1,t_2}\right]^{\Nx{\A}-\sx{1}-i}.
\label{FEB13_19}
\end{align}
\fi

Eq. (\ref{FEB13_19}) can be simplified by applying the Poisson approximation
to the binomial distribution and then assuming that $\Nx{\A}-\sx{1} \approx \Nx{\A}$, such
that (\ref{FEB13_19}) is no longer conditioned on $\sx{1}$ and becomes
\begin{equation}
\Pr\left(i \,\textnormal{arrive}\right) =
\frac{\left[\Nx{\A}\parrive{t_1,t_2}\right]^i
\EXP{-\Nx{\A}\parrive{t_1,t_2}}}{i!}.
\label{EQ13_04_16}
\end{equation}

Substituting all components back into (\ref{EQ13_05_15}) enables us to
write the two-observation joint probability distribution as
\ifOneCol
\begin{align}
& \Pr\left(\Nobst{t_1} = \sx{1}, \Nobst{t_2} = \sx{2}\right) =
\left[\Nx{\A}\Pobsx{t_1}\right]^{\sx{1}}
\EXP{-\Nx{\A}\left[\Pobsx{t_1}+\parrive{t_1,t_2}\right]} \nonumber \\
& \quad\times \sum\limits_{i = \max(0,\sx{1}-\sx{2})}^{\sx{1}}
\frac{\pleave{\deltObs}^i\left(1-\pleave{\deltObs}\right)^{\sx{1}-i}}{i!\left(\sx{1}-i\right)!}
\frac{\left[\Nx{\A}\parrive{t_1,t_2}\right]^{\sx{2}+i-\sx{1}}}
{\left(\sx{2}+i-\sx{1}\right)!}.
\label{EQ13_04_17}
\end{align}
\else
\begin{align}
& \Pr\left(\Nobst{t_1} = \sx{1}, \Nobst{t_2} = \sx{2}\right) = \nonumber \\
& \quad\left[\Nx{\A}\Pobsx{t_1}\right]^{\sx{1}}
\EXP{-\Nx{\A}\left[\Pobsx{t_1}+\parrive{t_1,t_2}\right]} \nonumber \\
& \quad\times \sum\limits_{i = \max(0,\sx{1}-\sx{2})}^{\sx{1}}
\frac{\pleave{\deltObs}^i\left(1-\pleave{\deltObs}\right)^{\sx{1}-i}}{i!\left(\sx{1}-i\right)!}
\nonumber \\
& \quad\times
\frac{\left[\Nx{\A}\parrive{t_1,t_2}\right]^{\sx{2}+i-\sx{1}}}
{\left(\sx{2}+i-\sx{1}\right)!}.
\label{EQ13_04_17}
\end{align}
\fi

Using (\ref{EQ13_04_17}) and (\ref{EQ13_04_03}),
we can evaluate (\ref{EQ13_04_18}) numerically for any pair of observation
times $t_1$ and $t_2$ and also compare with simulations that generate the
joint and marginal probability distributions.
We will see in Section~\ref{sec_num} that as $\deltObs$
increases, $\Ix{}{\Nobst{t_1};\Nobst{t_2}}$ decreases, for any value of $t_1$.

\section{Optimal Sequence Detection}
\label{sec_opt}

In this section, we derive the optimal sequence detector to give a lower
bound on the achievable bit error performance of any practical detector.
We present a modified version of the Viterbi algorithm to reduce the
computational complexity of optimal detection and facilitate its
implementation in simulations.

\subsection{Optimal Detector}

The optimal joint interval detection problem can be defined as follows. Let us
assume that the transmitter sequence $\dataSeq$ is $\B{}$ bits in length.
Within each bit
interval, the receiver makes $\M$ observations. The value of the $\smM$th
observation in the $j$th interval is labeled $\sx{j,\smM}$. We assume
that the sampling times within a single interval can be written as the
function $\gx{\smM}$, and we define a global time sampling function
$\tx{j,\smM} = (j-1)\T + \gx{\smM}$, where
$j = \{1,2,\ldots,\B{}\}, \smM = \{1,2,\ldots,\M\}$.
Let us briefly consider two examples of
$\gx{\smM}$. If a single observation is made when the maximum number
of molecules is expected, $\tmax$, as we considered
in \cite{RefWorks:662}, then
$\gx{\smM}$ has one value, $\gx{1} = \tmax$. If there are observations taken
at times separated by constant $\deltObs$, then $\gx{\smM} = \smM\deltObs$.

The optimal receiver decision rule is to select the sequence $\dataObs{j}$
that is most likely given the joint likelihood of all received samples, i.e.,
\begin{equation}
\label{FEB13_03}
\dataObs{j}\eqBar{j}{\{1,2,\ldots,\B{}\}} =
\argmax_{\data{j}, j = \{1,2,\ldots,\B{}\}} \Pr\left(\Nobs\right)
\end{equation}
where
\ifOneCol
\begin{equation}
\Pr\left(\Nobs\right) = \Pr\big(\Nobst{\tx{1,1}} = \sx{1,1},
\Nobst{\tx{1,2}} = \sx{1,2}, \ldots,
\Nobst{\tx{\B{},\M}} = \sx{\B{},\M} \mid \dataSeq \big).
\end{equation}
\else
\begin{align}
\Pr\left(\Nobs\right) = & \Pr\big(\Nobst{\tx{1,1}} = \sx{1,1}, \nonumber \\
& \Nobst{\tx{1,2}} = \sx{1,2}, \ldots, \nonumber \\
& \Nobst{\tx{\B{},\M}} = \sx{\B{},\M} \mid \dataSeq \big).
\end{align}
\fi

$\Pr(\Nobs)$ is the joint probability distribution function over all $\B{}\M$
observations, given a specified transmitter sequence $\dataSeq$.
Its form is readily tractable only if we assume that all
individual observations are independent of each other, i.e., if
\begin{equation}
\label{FEB13_21}
\Pr(\Nobs) = \prod_{j=1}^{\B{}}\prod_{\smM=1}^\M
\Pr\big(\Nobst{\tx{j,\smM}} = \sx{j,\smM} \mid \dataSeq)\big).
\end{equation}

If we apply our analysis in Section~\ref{sec_obs} and conclude that receiver
observations are independent, then we can use
(\ref{FEB13_21}) to determine the likelihood of a given $\dataSeq$.
However, this is still a problem with significant
complexity, especially for large $\B{}$, because we must determine the likelihood
of $2^\B{}$ possible $\dataSeq$s. The total complexity is only linear in the
number of samples $\M$ since a larger $\M$ only means that more terms are included
in the product in (\ref{FEB13_21}).

\subsection{Optimal Joint Detection Using Viterbi Algorithm}
We consider the Viterbi algorithm in order to reduce the computational complexity
of optimal joint detection and evaluate the corresponding probability of error
in simulations as a
benchmark comparison with simpler detection methods. The memory and computational
requirements of the Viterbi algorithm are likely still too high for effective
implementation in a molecular communication system.

The general Viterbi algorithm is described in detail in \cite[Ch. 5]{RefWorks:175}.
The algorithm builds a trellis diagram of states where the number of states depends
on the channel memory, and each path through the trellis represents one candidate
sequence.
Our modified implementation artificially
shortens the (explicit) channel memory and delays the decision of a given bit by the shortened
memory (as performed in methods such as delayed decision-feedback
sequence estimation in \cite{RefWorks:703}),
but we include the impact of \emph{all} prior \ISI\, on the current candidate
states. If the memory is increased to the actual channel memory, then
our method is equivalent to the regular Viterbi algorithm.

Theoretically, the channel memory of a diffusive environment is infinite; from
(\ref{JUN12_47_pobs}),
we see that $\Pobsx{t} \to 0$ only as $t \to \infty$.
However, in practice, there will be some finite number of bit intervals after
which the impact of a given transmission becomes negligible. While it is prudent
to include the impact of \ISI\, from all previous bit intervals, we limit the \emph{explicit}
channel memory to $\VAmem$ bit intervals. Thus, there will be $2^{\VAmem}$ trellis
states, where each state represents a candidate sequence for the previous $\VAmem$
bit intervals. Each state $\VAstate$ has two potential incoming paths, representing
the two possible transitions from previous states (each transition corresponds
to the possibility of the bit in the $(\VAmem+1)$th prior interval being $0$ or $1$).

We define $\VAdataObs{l}{i}, l = \{1,2,\ldots,j\}$ as the $l$th received bit according
to the $i$th path leading to state $\VAstate$. The \emph{current} log likelihood for
the $i$th path leading to state $\VAstate$ in the $j$th interval,
which is the likelihood associated
with only the observations in the \emph{most recent} bit interval and the candidate
sequence $\VAdataObs{l}{i}$,
is $\VAcurLL{j}{i}$ and is evaluated by
\ifOneCol
\begin{equation}
\VAcurLL{j}{i}  = \sum_{\smM=1}^\M\log\left(
\Pr\left(\Nobst{\tx{j,\smM}} = \sx{j,\smM}\right)
\right)
 = \sum_{\smM=1}^\M\log\left(
\frac{{\Nobsavgt}^{\sx{j,\smM}}\EXP{-\Nobsavgt}}{\sx{j,\smM}!}
\right),
\label{EQ13_05_01}
\end{equation}
\else
\begin{align}
\VAcurLL{j}{i} & = \sum_{\smM=1}^\M\log\left(
\Pr\left(\Nobst{\tx{j,\smM}} = \sx{j,\smM}\right)
\right) \nonumber \\
& = \sum_{\smM=1}^\M\log\left(
\frac{{\Nobsavgt}^{\sx{j,\smM}}\EXP{-\Nobsavgt}}{\sx{j,\smM}!}
\right),
\label{EQ13_05_01}
\end{align}
\fi
where we assume that the observations are independent and we apply the Poisson
approximation to the probability of observing a given number of information molecules.
We note that taking the logarithm has no influence on the optimality but facilitates
numerical evaluation.
The \emph{cumulative} log likelihood for the $\VAstate$th state in the
$j$th interval is the log
likelihood of the most likely path (and the corresponding bit sequence $\VAdataObs{l}{}$)
to reach the $\VAstate$th state, calculated
for \emph{all prior} bit intervals. We write this likelihood as $\VAcumLL{j}{}$
and it is found as
\begin{equation}
\VAcumLL{j}{} = \max \left(\VAcumLL{j-1}{1}+\VAcurLL{j}{1},
\VAcumLL{j-1}{2}+\VAcurLL{j}{2}\right),
\label{EQ13_05_02}
\end{equation}
where $\VAcumLL{j-1}{i}$ is the cumulative log likelihood of the
state prior to the $\VAstate$th state along the $i$th path.
For the $j$th bit interval,
$\VAcumLL{j}{}$ is the likelihood associated with the most likely
transmitter bit sequence to lead to the $\VAstate$th state.
Our modified Viterbi algorithm sequentially builds the trellis diagram by
determining $\VAcumLL{j}{}$ for every state in every bit interval of the transmission,
and keeping track of the candidate bit sequence $\VAdataObs{l}{}$
that led to $\VAcumLL{j}{}$.
At the end of the algorithm, the receiver makes its decision by selecting
the bit sequence $\VAdataObs{l}{}$ associated with the largest
value of $\VAcumLL{\B{}}{}$.

It is straightforward to see that we have reduced the complexity of optimal
joint detection by only needing to find the likelihood of $\B{}\,2^{\VAmem+1}$
sets of $\M$ observations, rather than the likelihood of $2^{\B{}}$ sets
of $\B{}\M$ observations. However, this is still a significant computational
burden on the receiver, so it is of interest to consider simpler detection methods.
Furthermore, the derivation of the expected bit error probability for
a maximum likelihood detector is not easily tractable, so we are restricted
to evaluating the bit error probability via simulation.

\section{Weighted Sum Detectors}
\label{sec_subopt}

In this section, we introduce the family of weighted sum detectors for diffusive
molecular communication. These detectors do not have the same memory and
computational requirements as maximum likelihood detectors, and we are
able to derive the expected bit error probability for a given transmitter sequence.

\subsection{Detector Design and Performance}

We assume that there is only sufficient
memory for the receiver to store the $\M$ observations made within a single bit interval,
and that it is incapable of evaluating likelihoods or considering the impact of
prior decisions. Under these limitations, an intuitive receiver design is to
add together the individual observations, with a weight
assigned to each observation, and then compare the sum with a pre-determined
decision threshold. This is the weighted sum detector and it is implemented
in neuron-neuron junctions; see \cite[Ch. 12]{RefWorks:587}.
The detector proposed in our previous work in \cite{RefWorks:662} is the simplest special case of a weighted sum detector, i.e., $\M=1$.
Under specific conditions, which we will discuss in Section~\ref{sec_opt_weights},
a particular selection of weights makes the performance of this detector
equivalent to the optimal detector described in Section~\ref{sec_opt}.

The decision rule of the weighted sum detector in the $j$th bit interval is
\begin{equation}
\dataObs{j} = \left\{
 \begin{array}{rl}
  1 & \text{if} \quad
  \sum_{\smM = 1}^{\M}\weight{\smM}\Nobst{\tx{j,\smM}} \ge \thresh,\\
  0 & \text{otherwise},
 \end{array} \right.
\label{FEB13_33}
\end{equation}
where $\weight{\smM}$ is the weight of the $\smM$th observation and $\thresh$ is the
binary decision threshold. For positive integer weights, we only need to consider
positive integer decision thresholds.

The method of calculation of the expected error probability
is dependent on the selection
of the weights. In a special case, if the weights are all equal, particularly if
$\weight{\smM} = 1\, \forall \smM$, and we assume that the value of each individual
observation is a Poisson random variable, then the weighted sum will also be
a Poisson random variable whose mean is the sum of the means of the individual
observations. Thus, we can immediately write the \CDF\, of the weighted sum in the
$j$th bit interval as
\ifOneCol
\begin{equation}
\Pr\left(\sum_{\smM = 1}^{\M}\Nobst{\tx{j,\smM}} < \thresh \right) =
\EXP{-\sum_{\smM = 1}^{\M}\Nobsavg{\tx{j,\smM}}}
\sum_{i=0}^{\thresh-1}
\frac{\left(\sum\limits_{\smM = 1}^{\M}\Nobsavg{\tx{j,\smM}}\right)^i}{i!}.
\label{EQ13_06_02}
\end{equation}
\else
\begin{align}
& \Pr\left(\sum_{\smM = 1}^{\M}\Nobst{\tx{j,\smM}} < \thresh\right) = \nonumber \\
& \qquad\EXP{-\sum_{\smM = 1}^{\M}\Nobsavg{\tx{j,\smM}}} \nonumber \\
& \qquad\times \sum_{i=0}^{\thresh-1}
\frac{\left(\sum\limits_{\smM = 1}^{\M}\Nobsavg{\tx{j,\smM}}\right)^i}{i!}.
\label{EQ13_06_02}
\end{align}
\fi

We note that, especially if $\M$
is large, relevant values of $\thresh$ may be very high, even if the expected
number of molecules counted in a single observation is low. Thus,
we might have difficulty in evaluating (\ref{EQ13_06_02})
numerically. We present two alternate methods to evaluate (\ref{EQ13_06_02}).
First, we can write the \CDF\, of a Poisson random variable $X$ with mean $\lambda$
in terms of Gamma functions, as \cite[Eq. (1)]{RefWorks:708}
\begin{equation}
\Pr\left(X < x\right) = \frac{\GamFcn{\lceil x\rceil,\lambda}}{\GamFcn{\lceil x\rceil}},
\label{EQ13_06_03}
\end{equation}
for $x > 0$, where the Gamma and incomplete Gamma functions are defined by
\cite[Eq. (8.310.1), Eq. (8.350.2)]{RefWorks:402}
\ifOneCol
\begin{equation}
\GamFcn{x} = \int_0^\infty \EXP{-a}a^{x-1}da, \quad
\GamFcn{x,\lambda} = \int_\lambda^\infty \EXP{-a}a^{x-1}da,
\label{EQ13_06_04}
\end{equation}
\else
\begin{align}
\GamFcn{x} & = \int_0^\infty \EXP{-a}a^{x-1}da, \nonumber \\
\GamFcn{x,\lambda} & = \int_\lambda^\infty \EXP{-a}a^{x-1}da,
\label{EQ13_06_04}
\end{align}
\fi
respectively. Second, the Gaussian approximation of the Poisson distribution
becomes more appropriate as the mean of the Poisson distribution increases.
Including a continuity correction (appropriate when approximating a discrete
random variable with a continuous random variable; see \cite[Ch. 6]{RefWorks:725}),
the \CDF\, has the form
\begin{equation}
\Pr\left(X < x\right) = \frac{1}{2}
\left[1 + \ERF{\frac{x - 0.5 - \lambda}{\sqrt{2\lambda}}}\right].
\label{EQ13_06_05}
\end{equation}

Now we consider the more general case, where we have positive non-equal weights.
Our analysis can then be summarized in the following theorem:

\begin{theorem}[Distribution of a weighted sum]
\label{theorem_weighted}
Given $\M$ Poisson random variables with means $\lambda_\smM$ and non-negative
weights $\weight{\smM}$, then the weighted sum
$X = \sum_{\smM=1}^\M\weight{\smM}X_\smM$ is in general not a Poisson
random variable, however
the weighted sum of Gaussian approximations of the individual variables
is a Gaussian random variable with mean $\sum_{\smM=1}^\M\weight{\smM}\lambda_\smM$
and variance $\sum_{\smM=1}^\M\weight{\smM}^2\lambda_\smM$.
\end{theorem}
\begin{IEEEproof}
The proof is straightforward to show using moment generating functions; see
\cite[Ch. 4]{RefWorks:725}. It can be shown that a sum of weighted independent
Poisson random
variables is also a Poisson random variable \emph{only if} the weights
$\weight{\smM}$ are all equal to $1$. However,
the Gaussian approximation of each $X_\smM$ gives a Gaussian random variable with
mean and variance $\lambda_\smM$, and it can be shown that \emph{any} weighted sum of
Gaussian random variables
is also a Gaussian random variable.
\end{IEEEproof}

Using Theorem~\ref{theorem_weighted}, we can immediately write the \CDF\, of random
variable $X = \sum_{\smM=1}^\M\weight{\smM}X_\smM$ as
\begin{equation}
\Pr\left(X < x\right) = \frac{1}{2}
\left[1 + \ERF{\frac{x - 0.5 - \sum_{\smM=1}^\M\weight{\smM}\lambda_\smM}
{\sqrt{2\sum_{\smM=1}^\M\weight{\smM}^2\lambda_\smM}}}\right].
\label{EQ13_06_12}
\end{equation}

In summary, for evaluation of the expected error probability when the
weights are equal, we can use the \CDF\,
(\ref{EQ13_06_02}), its equivalent form (\ref{EQ13_06_03}), or its
approximation (\ref{EQ13_06_05}) where
$\lambda = \sum_{\smM = 1}^{\M}\weight{\smM}\Nobsavg{\tx{j,\smM}}$.
When the weights are not equal, we must use (\ref{EQ13_06_12}) where
$\lambda_\smM = \Nobsavg{\tx{j,\smM}}$.

Given a particular sequence
$\dataSeq = \{\data{1},\ldots,\data{\B{}}\}$,
we can
write the probability of error of the $j$th bit,
$\Pe{j | \dataSeq}$, as
\ifOneCol
\begin{equation}
\Pe{j | \dataSeq} = \left\{
 \begin{array}{rl}
  \Pr\left(\sum_{\smM = 1}^{\M}\weight{\smM}\Nobst{\tx{j,\smM}} < \thresh \right)
  & \text{if} \;\data{j} = 1,
  \vspace*{2mm}\\
  \Pr\left(\sum_{\smM = 1}^{\M}\weight{\smM}\Nobst{\tx{j,\smM}} \ge \thresh \right)
  & \text{if} \;\data{j} = 0,
 \end{array} \right.
 \label{EQ13_06_15}
\end{equation}
\else
\begin{align}
& \Pe{j | \dataSeq} = \nonumber \\
& \quad\left\{
 \begin{array}{rl}
  \Pr\left(\sum_{\smM = 1}^{\M}\weight{\smM}\Nobst{\tx{j,\smM}} < \thresh \right)
  & \text{if} \;\data{j} = 1,
  \vspace*{2mm}\\
  \Pr\left(\sum_{\smM = 1}^{\M}\weight{\smM}\Nobst{\tx{j,\smM}} \ge \thresh \right)
  & \text{if} \;\data{j} = 0.
 \end{array} \right.
 \label{EQ13_06_15}
\end{align}
\fi

The true \emph{expected} error probability for the $j$th bit is found by evaluating
(\ref{EQ13_06_15}) for all possible transmitter sequences and scaling
each by the likelihood of that sequence occurring in a weighted sum
of all error probabilities, i.e.,
\begin{equation}
\Peavg{j} = \sum_{\dataSeq \in \dataSet}
\Pone^{\numX{1,j}{\dataSeq}}(1-\Pone)^{\numX{0,j}{\dataSeq}}\Pe{j | \dataSeq},
\end{equation}
where $\dataSet$ is the set of all $2^B$ possible sequences and
$\numX{i,j}{\dataSeq}$ is the number of occurrences of bit $i$ within the
first $j$ bits of the sequence $\dataSeq$. In practice, we can randomly generate
a large number (e.g., some hundreds or thousands) of sequences based on $\Pone$,
and we will find that it is sufficient to average the probability of error over
this subset of total possible sequences.

\subsection{Optimal Weights}
\label{sec_opt_weights}

We now consider optimizing the sample weights.
It would make sense to assign greater weight
to samples that are expected to measure more information molecules. We
make this claim more specific in the context of the matched filter.

Consider a signal $\hx{t}$. The impulse response of the matched filter to
signal $\hx{t}$ is $\hx{\T-t}$, where $\T$ is the signal interval;
see \cite[Ch. 5]{RefWorks:175}. The output of such a filter
at time $\T$ is then
\begin{equation}
\int\limits_0^{\T} \hx{\tau}\hx{\T-\T+\tau} d\tau =
\int\limits_0^{\T} \hx{\tau}^2 d\tau,
\label{EQ13_06_17}
\end{equation}
which is the continuous time equivalent of a weighted sum detector where the
sample at time $t$ is simply weighted by the expected value of the signal
of interest. Thus, we design a matched filter detector by
setting the sample weight $\weight{\smM}$ equal to the number of molecules
expected from the transmitter, $\Ntxtavg{\gx{\smM}}$.

The matched filter is optimal in the sense that it maximizes the signal-to-noise ratio, and it also minimizes the bit error probability if the desired signal
is corrupted by additive white Gaussian noise (\AWGN).
We generally cannot claim that the desired signal is corrupted by \AWGN, as that
would require a large expected number of molecules at all samples and an \AWGN\,
external noise source. However, if these conditions were satisfied, and $\T$ was
chosen to be sufficiently long to ignore the impact of \ISI, then the optimal
weighted sum detector would have weights $\weight{\smM} = \Nx{\A}\Pobsx{\gx{\smM}}$,
and it would be equivalent to the optimal sequence detector.

\subsection{Optimal Decision Threshold}

Our discussion of weighted sum detectors has not included the selection of the
decision threshold $\thresh$. Obviously, the performance of a particular detector
relies on the chosen value of $\thresh$. The derivation of the optimal $\thresh$
for a weighted sum detector is left for future work. In Section~\ref{sec_num},
when we discuss the bit error probability observed via simulation of a weighted
sum detector, we imply that the optimal $\thresh$ for the given environment
was found via numerical search.

\section{Numerical Results}
\label{sec_num}

In this section, we present simulation results to assess the performance
of the detectors described in this paper. Our simulations are
executed in a particle-based
stochastic framework where time is advanced in discrete steps $\Delta t$.
We note that if receiver observations are made at intervals $\deltObs$,
then $\deltObs$ must be a multiple of $\Delta t$.
For each time step, molecules undergo random motion over continuous space, and
the reaction rate constants are used to find the probabilities of reactions
(\ref{k1_mechanism})-(\ref{k2_mechanism}) occurring within a given step.
For further details, please refer to \cite{RefWorks:631, RefWorks:662}.
The inclusion of steady uniform flow as a constant displacement in addition
to random diffusion is straightforward, and
the impact of additive noise sources is added post-simulation to receiver
observations.

In order to focus on a comparison of the performance of the different detectors,
most physical
parameters remain constant throughout this section. The only environmental
parameters that we adjust are the quantity of additive noise, the distance $\x_0$ to the receiver, the degree of
steady uniform flow, and whether enzymes are present. For communication, we
vary the length of the bit interval and the frequency of sampling. In the following,
we refer to the \emph{base case} as numerical results found when
there is no additive noise, no flow, and no active enzymes present.

We model a fluid environment with a uniform viscosity of
$10^{-3}\,\textnormal{kg}\cdot\metre^{-1}\second^{-1}$ and at a temperature of
$25\,^{\circ}\mathrm{C}$ (i.e., the solvent is water). The transmitter emits impulses of $\Nemit = 5000$
$\A$ molecules to transmit a binary $1$ with probability $\Pone = 0.5$.
When enzymes are present, they have
a concentration of $84\,\mu\textnormal{M}$, and the values of $\kth{1}, \kth{-1}$,
and $\kth{2}$ are $2\times10^{-19}\, \frac{\metre^3}{\molecule\cdot\second}$,
$10^4\, \second^{-1}$, and $10^6\,\second^{-1}$, respectively. The radii of the $\A$,
$\En$, and $\EA$ molecules are $0.5\,$nm, $2.5\,$nm, and $3\,$nm, respectively
(i.e., using (\ref{JUN12_60}) we get
$\Dx{\A} = 4.365\times 10^{-10}\frac{\metre^2}{\second}$).
Unless otherwise noted, the
receiver is centered at $\x_0 = 300\,$nm and has a radius $\robs = 45\,$nm.
The simulation time step is $\Delta t = 0.5\,\mu\second$. Although the
environment is relatively small (the size of the receiver and its distance
from the transmitter are both significantly less
than the size of a bacterial cell)
with a low number of $\A$ molecules and high chemical reactivity,
the channel impulse response
scales to dimensionally homologous systems that have (for example) more molecules
and lower reactivities; see \cite{RefWorks:706}.
Our choice of parameters is made to ease the time required
for simulations.

\subsection{Sample Independence}

The design of the optimal sequence detector is based on the assumption that all observations
made at the receiver are independent of each other. Thus, we consider the mutual information
between receiver observations as a function of the time between them, when
the only emission of molecules by the transmitter is at time $t = 0$. In Fig.~\ref{fig1},
we show the evaluation of the mutual information for the base case when the
reference samples are taken at times $t_1 = \{10, 20, 50\}\,\mu\second$.
Similar simulation results are observed when flow is present or enzymes are added
and we omit these results for clarity.
For the expected
value curves, we evaluate (\ref{EQ13_04_18}) numerically as described in
Section~\ref{sec_indp}, whereas the simulation curves
are found by constructing the
joint and marginal \PDF s using $5\times10^5$ independent simulations.
Base $2$ is used in the logarithm so that mutual information is measured in bits.

\ifOneCol
\else
	\figOne{!tb}
\fi

In Fig.~\ref{fig1}, we see that the mutual information drops below $0.01$ bits within
$4\,\mu\second$ of all reference samples. The agreement
between the expected value curves and those generated via simulations is quite
good and only limited
by the accuracy of $\pleave{t}$ and $\parrive{t}$ (which critically rely on
the uniform concentration assumption).

\subsection{Detector Performance}

We now make a comparison of the average bit error probabilities obtainable via
the optimal sequence detector, the matched filter detector, and the
equal weight detector. We note that the bit error probability observed for the optimal
detector is found assuming independent samples via simulation only.
For the matched filter detector, the weights are
based on the number of molecules from the transmitter expected at the receiver, $\Ntxtavg{t}$,
due to a transmitter emission in the \emph{current} interval only (so that the weights are the
same for every bit interval; the adaptive assignment of weights, which is also
performed by neurons as described in \cite[Ch. 12]{RefWorks:587},
is left for future work).
For the equal weight detector, the expected error probability is
found using (\ref{EQ13_06_03}).
In every subsequent figure, we show the average bit
error probability as a function of the number of observations $\M$ made per
bit interval. For simplicity, observations are equally spaced within the interval,
such that $\gx{m} = m\T/\M$.

First, we consider the \ISI-free case by having the transmitter send a single bit.
We do this to assess whether the matched filter performs the same as the optimal
detector in the absence of \ISI. In
order for the detection to be meaningful, we add an independent Poisson noise source
with mean $50$ and impose a bit interval of $\T=200\,\mu\second$ for sampling (for reference,
the maximum value of $\Ntxtavg{t}$ under these conditions is $5.20$ molecules at
$34.36\,\mu\second$ after transmission).
The bit error
probability found via simulation is averaged over $2\times10^5$ transmissions,
and the expected error probability is found considering all bit sequences
since in this case there are only two (either $1$ or $0$). The results are presented
in Fig.~\ref{fig3}. We see that the bit error probability achievable with the
matched filter is equal to that of the maximum likelihood detector for any
value of $\M$. Thus, we claim that the performance of the matched filter is equivalent to the
optimal receiver design if there is no \ISI, even though the signal is corrupted
by non-\AWGN\, noise. Furthermore, the bit error probability achievable
with the equal weight detector is greater than those of the optimal detectors.
For example, the equal weight detector achieves a bit error probability of $0.03$
for $\M = 100$
samples, whereas the optimal detectors achieve a bit error probability of $0.017$.
Finally, the expected error probabilities evaluated for the matched filter and
equal weight detectors are very accurate when compared with the simulation results,
except when the error probability is below $0.01$ (where the deviation is primarily due to the increasing sample dependence).
Interestingly, there appears to be no degradation in bit error performance as $\deltObs$ goes to $0$ (i.e., as $\M$ increases), even for the maximum likelihood detector which was designed assuming independent samples.
However, for clarity, and in consideration of Fig.~\ref{fig1},
we separate samples
in subsequent simulations by at least $5\,\mu\second$.

\ifOneCol
\else
	\figThree{!tb}
\fi

In the subsequent figures, we consider transmitter sequences of $100$ consecutive bits,
and the error probabilities shown are averaged over all intervals and independent
realizations. The expected error probabilities are evaluated by taking the average
over $1000$ random bit sequences. For optimal joint detection we limit the
explicit channel memory to $\VAmem = 2$ bit intervals as a compromise between
computational complexity and observable error probability.

In Fig.~\ref{fig4}, we study the impact of adding noise to the base case when
the bit interval is $\T=200\,\mu\second$.
Due to \ISI, the bit error probability achievable
with the matched filter is not as low as that achievable with the optimal sequence
detector. The disparity reaches orders of magnitudes as $\M$ increases (about
$2$ orders of magnitude in the noiseless case when $\M = 20$).
For all detectors, the error probability rises as additive noise is introduced.
However, all detectors are able to achieve
a probability of error below $0.01$ as more samples are taken, even when
$\Nnoisetavg{t} = 0.5$.

\ifOneCol
\else
	\figFour{!tb}
\fi 

In Fig.~\ref{fig3_5}, we consider the impact of propagation distance on receiver performance when the bit interval is $\T=200\,\mu\second$ by varying $\x_0$ from $250\,$nm to $500\,$nm while keeping all other transmission parameters constant. Noise is not added. We see that all detectors are very sensitive to the propagation distance; the bit error probability varies over many orders of magnitude, even though the distances vary at most by a factor of 2. As the distance increases, fewer molecules reach the receiver (i.e., there is less received signal energy) and it takes more time for those molecules to arrive (i.e., the channel impulse response is longer and there is \emph{relatively} more \ISI, although the optimal sequence detector is more robust to this effect). The performance degradation at longer distances can be mitigated by methods such as increasing the bit interval time, changing the number of molecules released by the transmitter, or adding enzymes to the propagation environment.

\ifOneCol
\else
\figThreePointFive{!tb}
\fi

In Figs.~\ref{fig5} and \ref{fig6}, we consider the impact of the presence of
enzymes in the propagation environment in comparison with the base case
when the bit interval is $\T=100\,\mu\second$, i.e., shorter than that in Figs.~\ref{fig4} and \ref{fig3_5}.
In both Figs.~\ref{fig5} and \ref{fig6}, the expected error probabilities when enzymes
are present are not as accurate as in the base case.
This is because we assumed that
the concentration expected at the receiver as given in (\ref{JUN12_47}) is exact,
when in fact it is a lower bound (we used $\kth{1}$ for $\kth{}$).
As the expected error probability decreases, it becomes relatively more sensitive to
the accuracy of this bound.

\ifOneCol
\else
	\figFive{!tb}
\fi

In Fig.~\ref{fig5}, we consider no additive noise sources.
The bit interval is so short that, without enzymes, the bit error probability
observed by the weighted sum detectors reaches a floor of about $0.06$. Orders
of magnitude of improvement in bit error probability are observed when enzymes
are added to degrade the $\A$ molecules and mitigate the impact of
\ISI; both weighted sum detectors
give a bit error probability
below $0.005$ for $\M \ge 10$, and performance is within an order of magnitude
of the maximum likelihood detector. Interestingly, the equal weight detector
outperforms the matched filter detector for $\M < 10$, emphasizing that
the matched filter does not necessarily optimize the probability of error
when the noise is not \AWGN\, and \ISI\, is present (though the Gaussian
approximation of the noise improves as $\M$ increases).
There is also an improvement in the performance of the maximum likelihood detector
for the range of $\M$ shown, even though the enzymes are effectively
destroying signal energy that could have been used to help with joint detection
(the maximum likelihood detector performs better without enzymes when $\M > 15$,
but this is omitted from Fig.~\ref{fig5} to maintain clarity). The reason
for this improvement is the chosen value of $\VAmem = 2$; the actual channel memory
is much longer without enzymes.

In Fig.~\ref{fig6}, we include an additive noise source. Since
we do not model the actual location of the noise source, we cannot predict how much
of this noise will be broken down by enzymes before it reaches the receiver. Intuitively,
the enzymes will be able to more easily degrade the molecules emitted by the noise
source before they are observed by the receiver if the noise source is placed further away.
For a fair comparison, we consider a noise
source that is placed at a ``moderate'' distance from the receiver, such that
the receiver observes noise with mean $1$ in the base case and with mean $0.5$
when enzymes are present. The optimal detector can now clearly benefit from
the presence of enzymes for all values of $\M$, since the enzymes
break down noise molecules in addition to
those emitted by the transmitter. The improvement in bit error probability of the optimal
detector is about $20\,\%$ when enzymes are present, for all values of $\M$.
Of course, the error probabilities observed by all
detectors either with or without enzymes
are worse than those observed in the no-noise case in Fig.~\ref{fig5}.

\ifOneCol
\else
	\figSix{!tb}
\fi

In Fig.~\ref{fig7}, we consider the impact of flow when there is an additive noise
source with mean value $1$ (we assume that the amount of additive
noise observed is independent of the flow, but of course in practice that would depend
on the nature of the noise source). We set the bit interval to $\T=100\,\mu\second$,
the same as in Fig.~\ref{fig6}, so we can make comparisons with the detectors
in Fig.~\ref{fig6}. We plot the average error probability for three
different flows: $\vx{\x} = 0.003\,\metre/\second$ (i.e., towards the receiver
from the transmitter), $\vx{\x} = -0.001\,\metre/\second$,
and $\vx{\y} = 0.003\,\metre/\second$ (i.e., perpendicular to the
line between transmitter and receiver). The flow magnitudes are chosen so that
they affect but do not dominate the channel impulse response (the corresponding
Peclet numbers, which describe the dominance of convection versus diffusion
and are found here as $\frac{\x_0\vx{}}{\Dx{\A}}$, are $2.06$ for
$\vx{y}$ and the positive $\vx{x}$; see \cite[Ch. 5]{RefWorks:587}). Of course, for a given flow, the random diffusion of each molecule in every time step is added to the constant displacement due to flow.

\ifOneCol
\else
	\figSeven{!tb}
\fi

When $\vx{\x}$ is positive, the observed error probabilities in Fig.~\ref{fig7}
are much better than in the corresponding no-enzyme case in Fig.~\ref{fig6},
and also better than when enzymes are present, as the flow
both increases the strength of the observed signal and mitigates \ISI.
The expected performance of the weighted sum detectors is not as
accurate for positive $\vx{\x}$ because the uniform concentration assumption
at the receiver has even less validity with this flow.
Perhaps more
interestingly, we see that the observed error probabilities with the
weighted sum detectors are better than the no-enzyme case in Fig.~\ref{fig6} when
$\vx{\x}$ is negative (although
the magnitude of negative flow that we consider is less than that of the
positive flow, such that we are still able to observe $\A$ molecules at the receiver).
Furthermore, all detectors perform better than the no-enzyme case when the direction
of flow is perpendicular to the direction of information transmission. These ``disruptive'' flows,
which are not in the direction of information transmission, do
not prevent the ability to communicate, and in these cases improve
transmission. Thus, it is possible to
consider bi-directional transmission in flowing environments, independent of the
direction of the flow.
We conduct further study on the impact of steady uniform flows with a broader range of Peclet numbers in \cite{RefWorks:752}.

\section{Conclusion}
\label{sec_concl}

In this paper, we studied both optimal and suboptimal detectors for an ideal
receiver in a diffusive molecular communication environment. Our physical
model is a  general one that can include steady uniform flow in any arbitrary
direction, sources of information molecules in addition to the transmitter,
and enzymes in the propagation environment
to degrade the information molecules.
We derived the mutual information between receiver observations
to show how often independent observations can be made.
Furthermore, we derived
the maximum likelihood sequence detector to provide a lower bound on
the achievable bit error probability. We also designed weighted sum detectors
as a family of more practical detectors, where the optimal selection of weights
under corruption by \AWGN\,
is the matched filter and it performs very close to the optimal detector in
the absence of \ISI, even if the additive noise is not Gaussian.
Simpler weighted sum detectors, with either equal weights
or fewer samples per bit interval, offer an easier implementation
at the cost of higher error probability. We showed that having enzymes present
enables high throughput without relying on the complexity of
the optimal detector. We also showed that communication using
weighted sum detectors can be improved
in flowing environments, for any steady flow direction, so long
as the velocity of the flow is not too high.

\appendix
\label{app_pstay}

The integration in (\ref{MAR13_03}) to prove Theorem~\ref{theorem_pstay}
can be divided into three parts,
which we write as
\begin{align}
\label{MAR13_06}
& \frac{3}{2\robs^3}\int\limits_0^{\robs} r^2 \left(
\ERF{\frac{\robs-r}{2\sqrt{\Dx{\A}\deltObs}}} +
\ERF{\frac{\robs+r}{2\sqrt{\Dx{\A}\deltObs}}}\right) dr, \\
\label{MAR13_07}
& -\frac{3}{\robs^3}\sqrt{\frac{\Dx{\A}\deltObs}{\pi}}\int\limits_0^{\robs}
r\EXP{-\frac{(\robs-r)^2}{4\Dx{\A}\deltObs}} dr, \\
\label{MAR13_08}
& \frac{3}{\robs^3}\sqrt{\frac{\Dx{\A}\deltObs}{\pi}}\int\limits_0^{\robs}
r\EXP{-\frac{(\robs+r)^2}{4\Dx{\A}\deltObs}} dr.
\end{align}

We begin with (\ref{MAR13_07}). It is straightforward to show via the
substitution $x = \robs-r$ and the definition of the error function in
(\ref{APR12_32}) that (\ref{MAR13_07}) evaluates to
\begin{equation}
\label{MAR13_12}
-\frac{3\Dx{\A}\deltObs}{\robs^2}\ERF{\frac{\robs}{2\sqrt{\Dx{\A}\deltObs}}}
- \frac{6(\Dx{\A}\deltObs)^\frac{3}{2}}{\robs^3\sqrt{\pi}}
\!\left[\EXP{\frac{-\robs^2}{4\Dx{\A}\deltObs}}-1\right]\!.
\end{equation}

Similarly, (\ref{MAR13_08}) evaluates to
\ifOneCol
\begin{equation}
\label{MAR13_14}
\frac{6(\Dx{\A}\deltObs)^\frac{3}{2}}{\robs^3\sqrt{\pi}}
\left[\EXP{\frac{-\robs^2}{4\Dx{\A}\deltObs}}-\EXP{\frac{-\robs^2}{\Dx{\A}\deltObs}}\right]
+ \frac{3\Dx{\A}\deltObs}{\robs^2}\left[\ERF{\frac{\robs}{2\sqrt{\Dx{\A}\deltObs}}} -
\ERF{\frac{\robs}{\sqrt{\Dx{\A}\deltObs}}}\right].
\end{equation}
\else
\begin{multline}
\label{MAR13_14}
\frac{6(\Dx{\A}\deltObs)^\frac{3}{2}}{\robs^3\sqrt{\pi}}
\left[\EXP{\frac{-\robs^2}{4\Dx{\A}\deltObs}}-\EXP{\frac{-\robs^2}{\Dx{\A}\deltObs}}\right] \\
+ \frac{3\Dx{\A}\deltObs}{\robs^2}\left[\ERF{\frac{\robs}{2\sqrt{\Dx{\A}\deltObs}}} -
\ERF{\frac{\robs}{\sqrt{\Dx{\A}\deltObs}}}\right].
\end{multline}
\fi

To solve (\ref{MAR13_06}), we first apply the substitutions
$\y = (\robs \pm r)/(\Dx{\A}\deltObs)$ to integrals containing
the first and second error functions, respectively. This enables us to
rewrite (\ref{MAR13_06}) with a single error function, as
\begin{equation}
\label{MAR13_19}
\frac{3}{2\robs^3}\int\limits_0^{\frac{\robs}{\sqrt{\Dx{\A}\deltObs}}}
\left(\robs - 2\y\sqrt{\Dx{\A}\deltObs}\right)^2\ERF{\y} d\y.
\end{equation}

Evaluating (\ref{MAR13_19}) requires the solution of three integrals, being
the product of $\ERF{\y}$ with increasing powers of $\y$. Beginning with
the base case, from \cite[Eq. (5.41)]{RefWorks:402} we have
\begin{equation}
\label{MAR13_24}
\int\ERF{\y} d\y = \y\ERF{\y} + \frac{1}{\sqrt{\pi}}\EXP{-\y^2}.
\end{equation}

All terms in (\ref{MAR13_19}) can be evaluated using (\ref{MAR13_24}) and
integration by parts, such that (\ref{MAR13_06}) evaluates to
\ifOneCol
\begin{equation}
\left[\frac{3\Dx{\A}\deltObs}{\robs^2} + 1\right]\ERF{\frac{\robs}{\sqrt{\Dx{\A}\deltObs}}} +
\frac{1}{\robs}\sqrt{\frac{\Dx{\A}\deltObs}{\pi}}\left[\left(1 + \frac{4\Dx{\A}\deltObs}{\robs^2}\right)
\EXP{\frac{-\robs^2}{\Dx{\A}\deltObs}} -
3 + \frac{4\Dx{\A}\deltObs}{\robs^2}\right].
\label{MAR13_30}
\end{equation}
\else
\begin{multline}
\!\!\!\frac{1}{\robs}\sqrt{\frac{\Dx{\A}\deltObs}{\pi}}
\left[\!\left(1 + \frac{4\Dx{\A}\deltObs}{\robs^2}\right)
\EXP{\frac{-\robs^2}{\Dx{\A}\deltObs}} -
3 + \frac{4\Dx{\A}\deltObs}{\robs^2}\right] \\
 + \left[\frac{3\Dx{\A}\deltObs}{\robs^2} + 1\right]\ERF{\frac{\robs}{\sqrt{\Dx{\A}\deltObs}}}.
\label{MAR13_30}
\end{multline}
\fi

It is straightforward to combine (\ref{MAR13_12}), (\ref{MAR13_14}), and
(\ref{MAR13_30}) to arrive at (\ref{MAR13_31}).

\bibliography{../references/nano_ref}

\begin{IEEEbiography}[{\includegraphics[width=1in,height=1.25in,
		clip,keepaspectratio]{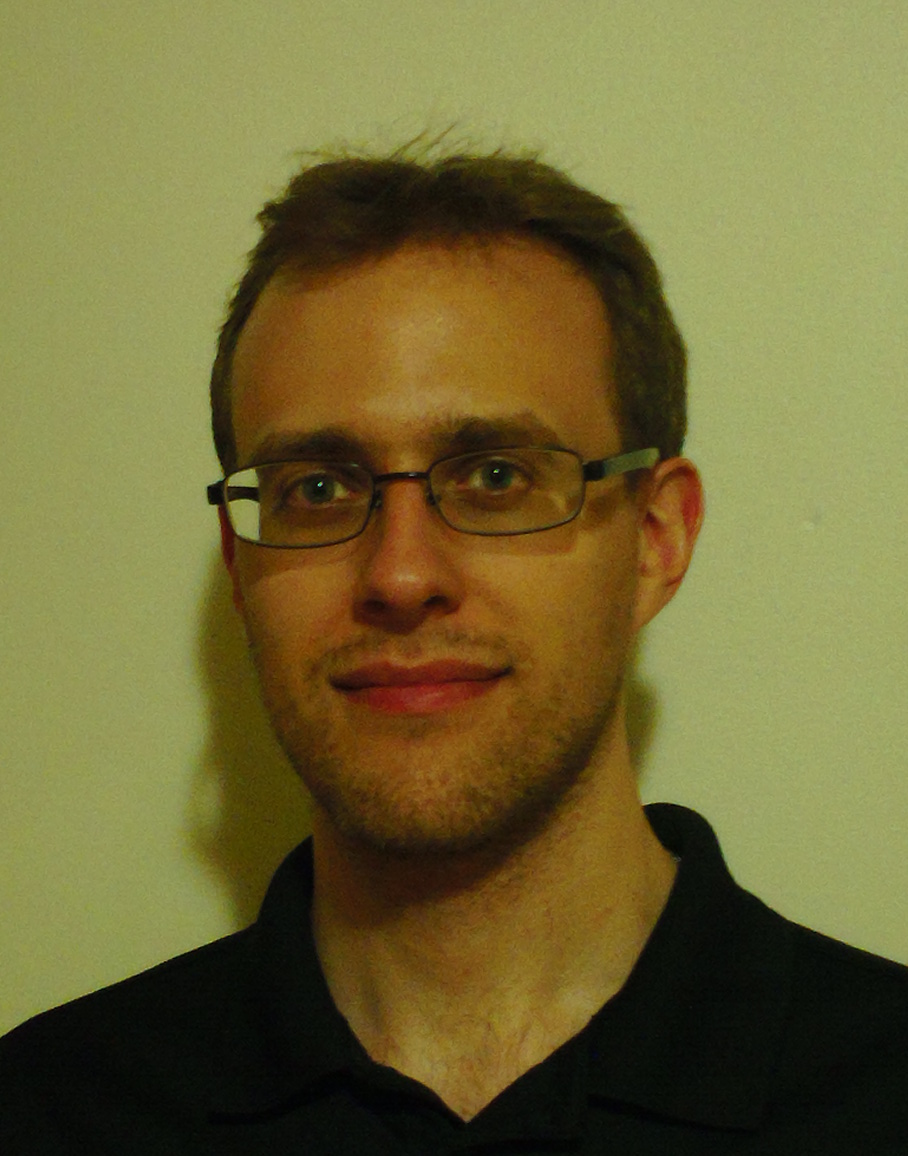}}]{Adam Noel}
	(S'09) received the B.Eng. degree from Memorial University in 2009 and the
	M.A.Sc.
	degree from the University of British Columbia (UBC) in 2011, both in
	electrical engineering.
	He is now a Ph.D. candidate in electrical
	engineering at UBC, and in 2013 was a visiting researcher
	at the Institute for Digital Communications,
	Friedrich-Alexander-Universit\"{a}t Erlangen-N\"{u}rnberg. His research
	interests include wireless communications and how traditional communication
	theory applies to molecular communication.
\end{IEEEbiography}

\begin{IEEEbiography}[{\includegraphics[width=1in,height=1.25in,
		clip,keepaspectratio]{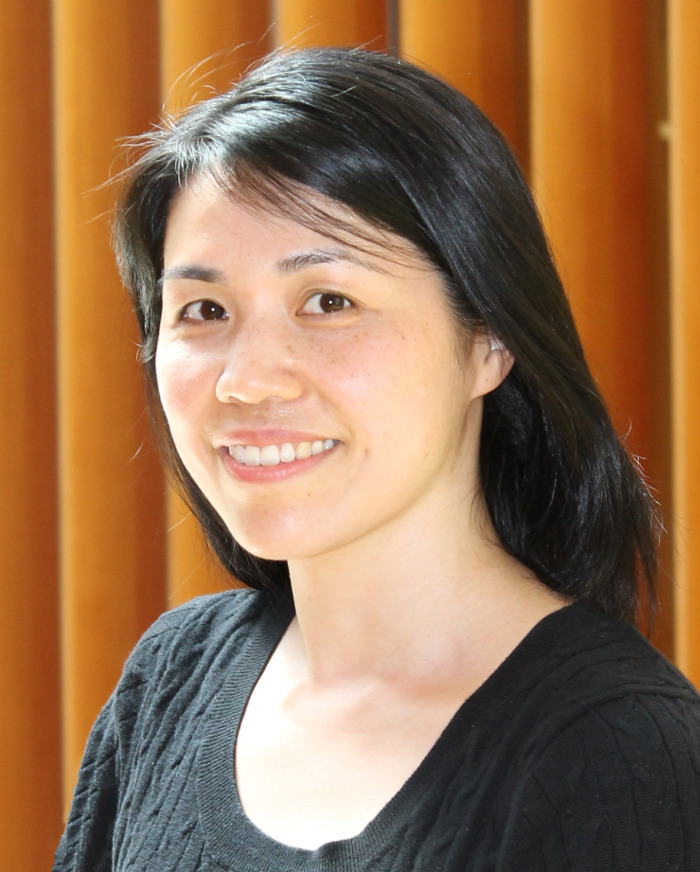}}]{Karen C. Cheung}
	received the B.S. and Ph.D. degrees in bioengineering from the University of
	California, Berkeley, in 1998 and 2002, respectively. From 2002 to 2005, she was
	a postdoctoral researcher at the Ecole Polytechnique Fédérale de Lausanne,
	Lausanne, Switzerland. She is now at the University of British Columbia,
	Vancouver, BC, Canada. Her research interests include lab-on-a-chip systems for
	cell culture and characterization, inkjet printing for tissue engineering, and
	implantable neural interfaces.
\end{IEEEbiography}

\begin{IEEEbiography}[{\includegraphics[width=1in,height=1.25in,
		clip,keepaspectratio]{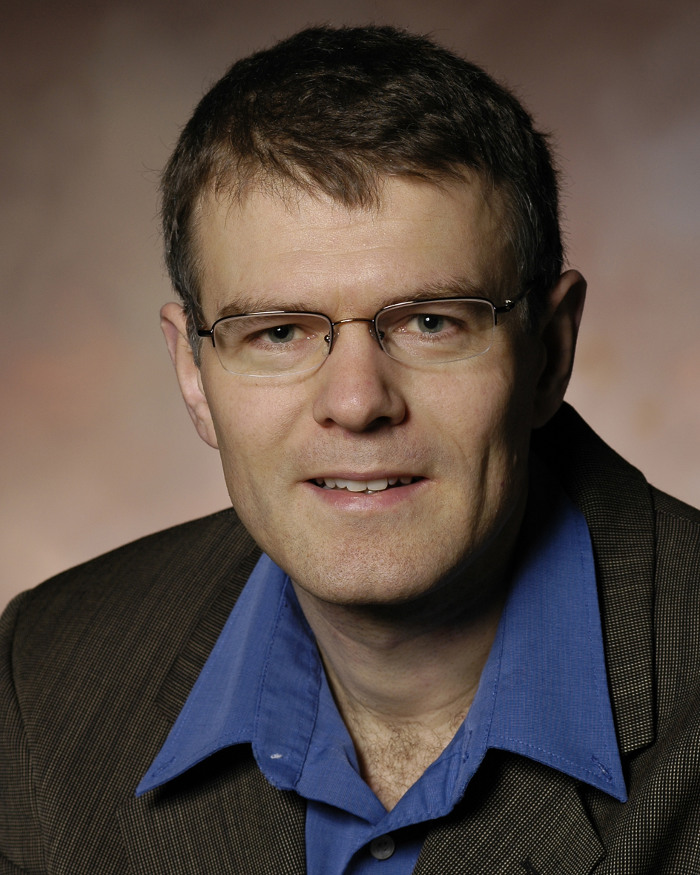}}]{Robert Schober}
	(S'98, M'01, SM'08, F'10)
	received the Diplom (Univ.) and the Ph.D. degrees in electrical engineering from
	the University of Erlangen-Nuremberg in 1997 and 2000, respectively.
	Since May 2002 he has been with the University of British Columbia (UBC),
	Vancouver, Canada, where he is now a Full Professor.
	Since January 2012 he is an Alexander
	von Humboldt Professor and the Chair for Digital Communication at the Friedrich
	Alexander University (FAU), Erlangen, Germany. His research interests fall into
	the broad areas of Communication Theory, Wireless Communications, and
	Statistical Signal Processing.
	He is currently the
	Editor-in-Chief of the IEEE Transactions on Communications.
	
\end{IEEEbiography}

\ifOneCol
	\newpage
	
		\figOne{H}
	
	\newpage
	
		\figThree{H}
	
	\newpage
	
		\figFour{H}
		
		\newpage
		
		\figThreePointFive{H}
	
	\newpage
	
		\figFive{H}
	
	\newpage
	
		\figSix{H}
	
	\newpage
	
		\figSeven{H}
	
	\newpage
	
	\section*{Figure captions}
	
	Fig. 1 \quad The mutual information in bits measured as a function
	of $t_o$.
	
	Fig. 2 \quad Expected error probability as a function of $\M$
	when there is no \ISI,
	$\Nnoisetavg{t} = 2$, and $\T = 500\,\mu\second$. The performance
	of the matched filter detector is equivalent to that of the
	maximum likelihood detector.
	
	Fig. 3 \quad Receiver error probability as a function of $\M$
	when \ISI\, is included, $\T = 200\,\mu\second$,
	and $\Nnoisetavg{t} = 0$ or $0.5$.
	
	Fig.~4 \quad Receiver error probability as a function of $\M$
	when \ISI\, is included, $\T = 200\,\mu\second$,
	and the distance $\x_0$ to the receiver is varied.
	
	Fig. 5 \quad Receiver error probability as a function of $\M$
	when \ISI\, is included, $\T = 100\,\mu\second$,
	and enzymes are added to mitigate the impact of \ISI.
	
	Fig. 6 \quad Receiver error probability as a function of $\M$
	when \ISI\, is included, $\T = 100\,\mu\second$,
	enzymes are added to mitigate the impact of \ISI, and an additive noise
	source is present ($\Nnoisetavg{t} = 1$ without enzymes and
	$\Nnoisetavg{t} = 0.5$ when enzymes are present).
	
	Fig. 7 \quad Receiver error probability as a function of$\M$
	when \ISI\, is included, $\T = 100\,\mu\second$,
	$\Nnoisetavg{t} = 1$, and different degrees of flow are present.
\fi

\end{document}